\documentclass[a4paper,twoside,british]{elsarticle}
\usepackage[T1]{fontenc}
\usepackage[latin9]{inputenc}
\usepackage{array}
\usepackage{amsmath}
\usepackage{graphicx}
\usepackage{setspace}

\makeatletter

\providecommand{\tabularnewline}{\\}
\newcommand{\lyxdot}{.}

\journal{ }

\usepackage{lineno}

\makeatother

\usepackage{babel}
\begin{document}

\begin{frontmatter}{}

\title{Numerical study of wet plastic particle separation using a coupled
DEM-SPH method}

\author[RUB]{Darius~Markauskas\corref{cor1}}

\ead{markauskas@leat.rub.de, Tel: +49 (0)234 / 32-27281, Fax: +49 (0)234
/ 32-14227 }

\author[TUB]{Harald~Kruggel-Emden}

\author[RUB]{Viktor~Scherer }

\cortext[cor1]{Corresponding author}

\address[RUB]{Ruhr-University Bochum, Universitaetsstrasse 150, D-44780 Bochum,
Germany}

\address[TUB]{Technical University of Berlin, Ernst-Reuter Platz 1, D-10587 Berlin,
Germany}
\begin{abstract}
The separation of different kind of plastic particles is required
in the process of waste recycling. For the separation drum processes
passed through by a liquid are applicable. Thereby the separation
is based on the principle that particles either sink or float in a
liquid depending on their densities. In this study the aforementioned
process is numerically analysed for the separation of polyethylene
terephthalate (PET) from polypropylene (PP) particles. The Discrete
Element Method coupled with the Smoothed Particle Hydrodynamics method
(DEM-SPH) is used for modelling purposes. The employment of the SPH
for the modelling of the liquid let us exploit the strong side of
this mesh-less method, namely, the relative easiness to model large
movements of the fluid together with free surfaces and moving boundaries.
The used theoretical model is presented and validation tests are performed,
where a dam-break problem is considered as an example. Simulations
of the plastic particle separation in the rotating drum are performed
thereafter. The influence of the different operational and design
parameters, such as the rotational velocity, the feed rate, the number
of lifters etc., on the resultant purity of the plastic is estimated.
It is expected that in the future the performed analysis will allow
to optimise drum separation processes. 
\end{abstract}
\begin{keyword}
solid-liquid flow \sep wet particle separation \sep fluid-particle
interaction \sep discrete element method \sep smoothed particle
hydrodynamics

\end{keyword}

\end{frontmatter}{}

\section{Introduction}

Mechanical plastic recycling is currently one of the weakest steps
in the recycling system, because only a low percentage of plastic
is reused compared to the amount of recovered metal, glass and waste
paper {[}\citealp{Dodbiba2002}{]}. As the raw mixture of plastic
waste usually includes various kinds of plastics (e.g. Acrylonitrile-butadiene-styrene
(ABS), Polyethylene terephthalate (PET), Polystyrene (PS), Polyethylene
(PE), Polypropylene (PP), Polyvinyl chloride (PVC)), the separation
process should classify waste into a number of reclaimable plastic
fractions, so as to meet the requirements for the purity and cleanliness
of a polymer type that are needed in a high quality plastic recycling
process {[}\citealp{Delgado2005}{]}. 

\begin{figure}
\begin{centering}
\includegraphics[width=14cm]{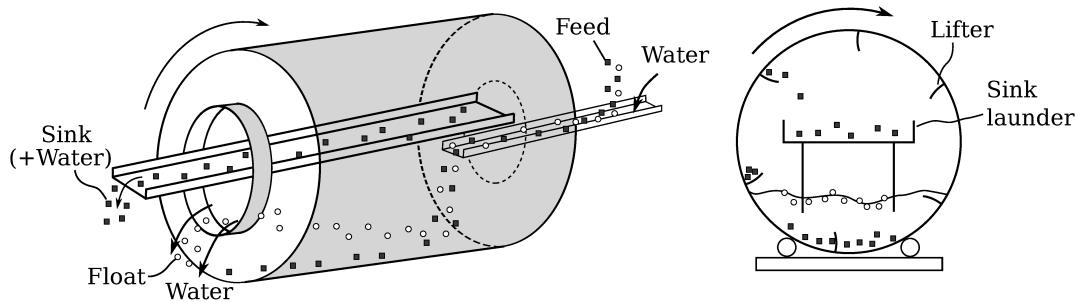}
\par\end{centering}
\caption{Scheme of a drum separator \label{fig:Scheme-Drum}}
\end{figure}

Mechanical plastic recycling processes can be divided into wet and
dry separation techniques. Among the wet processes drum separators
based on the float-sink principle {[}\citealp{Brandrup1996}{]} are
widely used for separating granular materials (Fig.\,\ref{fig:Scheme-Drum}).
They are based on the fact that grains either sink or float in liquids
depending on their densities. Before the separation process, the plastic
wastes are shredded into small particle-like entities. The resulting
plastic particle mixture is fed together with water into a rotating
drum. The sinking particles are lifted onto a sink launder and are
then removed from the drum, while the floating particles are discharged
out of the vessel together with the water. To shift the cut point
to a different density, it may be necessary to vary the density of
the separating liquid. This is done either by dissolving materials
of lower or greater density than water, for example alcohols or salts,
or by suspending fine grained solids of greater density than water.
The latter separating liquid is also known as a heavy medium and is
extensively employed in mineral processing industry {[}\citealp{Sripriya2006}{]}.

The successful separation of binary plastic mixtures in rotating drums
was studied experimentally by Dodbiba et al. {[}\citealp{Dodbiba2002}{]}.
A review of varying separation technologies and their efficiencies
was compiled by the same authors {[}\citealp{Dodbiba2004}{]}. For
multi-component plastic mixtures a three stage processes {[}\citealp{Pongst2008}{]}
was proposed which results in good recovery rates for most kinds of
plastic. Difficulties arise in float-sink separation if density differences
between plastic fractions become low or in case of elevated feed rates
{[}\citealp{Menad2013}{]}.

Potentially, the numerical modelling could contribute for the improvement
of wet separation of particles. Thereby for modelling of plastic particles,
a Discrete Element Method (DEM) can be used. DEM was first proposed
by Cundall and Strack in 1979 and since then is widely used in many
areas of powder technology and mechanical process engineering {[}\citealp{Zhu2008,Zhong2016}{]}.
In the DEM, solid particles are considered to be separate entities,
interacting with each other by contacts mostly. This discrete character
of the method allows a reduced set of constitutive assumptions to
be used as compared to continuum approaches.

By modelling of a mixed fluid-particle system, a coupling between
solid particles and fluid is required. In the coupled system, the
fluid phase can be modelled at the sub-particle level such that momentum
exchange between fluid and particle is resolved in detail {[}\citealp{Han2007}{]},
or using local averaging techniques {[}\citealp{Anderson67}{]}. The
simulations at the sub-particle level allow a detailed analysis of
interaction forces, that act between the fluid and solid particles,
and, therefore, can be used, e.g. for determination of drag correlations
{[}\citealp{Kruggel2016}{]}. However, such simulations are computationally
very expensive and, therefore, they usually are limited to a small
number of solid particles {[}\citealp{Wang2010,Zhao2013}{]}. The
use of local averaging techniques is computationally more efficient
and allows simulations of much larger particle systems, while preserving
the discrete characteristics of the particle flow.

For the modelling of the fluid flow, Navier-Stokes equations are usually
solved using mesh-based methods, like Finite Volume or Finite Element
Methods {[}\citealp{Ferziger2002}{]}. However, the application of
mesh based methods for modelling of complex geometries and free surfaces,
can be a challenging task {[}\citealp{Gao2009}{]}. To elevate these
difficulties a Smoothed Particle Hydrodynamics (SPH) method can be
used. The SPH, originally proposed by Gingold and Monaghan {[}\citealp{Gingold1977}{]}
and Lucy {[}\citealp{Lucy1977}{]}, is a mesh-less Lagrangian technique,
which proved to be a suitable tool for modelling of fluids in such
areas as marine {[}\citealp{Shibata2007}{]}, extrusion {[}\citealp{Prakash2015}{]},
geophysical {[}\citealp{Sivanesapillai2014}{]} or costal {[}\citealp{Gotoh2006}{]}
engineering. The major strength of this method is its mesh-less character,
which makes the method very flexible and enables the simulation of
engineering problems, that might be difficult to capture by conventional
mesh-based methods.

Recently, the two-way coupling between DEM and SPH based on a local
averaging technique was proposed by Gao and Herbst {[}\citealp{Gao2009}{]},
Sun et al. {[}\citealp{Sun2013}{]} and Robinson et al. {[}\citealp{Robinson2014}{]}.
The successful application of the two-way coupled DEM-SPH to slurry
flow, abrasive wear and magnetorheological fluids was demonstrated
by Cleary {[}\citealp{Cleary2015}{]}, Beck \& Eberhard {[}\citealp{Beck2015}{]}
and Lagger et al. {[}\citealp{Lagger2015}{]} respectively. A detailed
analysis of the sedimentation of one particle and a porous block presented
in {[}\citealp{Robinson2014}{]} and a comparative study on coupling
the DEM with mesh-based methods and DEM coupling with the mesh-less
SPH method reported by Markauskas et al. {[}\citealp{Markauskas2017}{]}
allow to conclude, that DEM-SPH is an appropriate and promising tool
for modelling particle-laden fluid systems. 

In the current investigation the wet separation of plastic particles
in a rotating drum is analysed numerically. The two-way coupled DEM-SPH
method is used for this purpose. In our earlier work {[}\citealp{Markauskas2017}{]}
the study on the coupling was presented, while in the current investigation,
the earlier developed famework is applied to a real engineering problem.
The flexibility of the DEM-SPH method enables us to simulate rapid
movements of the particle-laden fluid with free surfaces and moving
walls, and allows to analyse the influence of various operational
and design parameters on the separation process. To our best knowledge,
the wet particle separation in a rotating drum is analysed numerically
for the first time. 

\section{Governing equations\label{sec:Governing-eq}}

\subsection{Governing equations of the solid phase}

The discrete element method is used for modelling of the solid particles.
The motion of each particle is governed by Newton's second law:

\begin{equation}
m_{i}\frac{d\mathbf{u}_{i}}{dt}=\mathbf{F}_{i}^{c}+\mathbf{F}_{i}^{g}+\mathbf{F}_{i}^{int},\label{eq:sp_newton}
\end{equation}
where $\mathbf{u}_{i}$ is the solid particle velocity, $\mathbf{F}_{i}^{c}$
is the contact force, $\mathbf{F}_{i}^{g}$ is the gravity force.
$\mathbf{F}_{i}^{int}$ is the interaction force acting between solid
and fluid phase. The details how this force is calculated is given
in Section \ref{sec:Interaction}. The contact force for particle
$\mathcal{P}_{i}$ is obtained as a sum of all contact forces between
$\mathcal{P}_{i}$ and particles in contact $\mathcal{P}_{j}$:

\begin{equation}
\mathbf{F}_{i}^{c}=\sum_{j=1}^{n}\mathbf{F}_{ij}^{c}\:,\label{eq:ContactSum}
\end{equation}
where $n$ is the number of contacts. For the calculation of the contact
force in the normal direction a Hertz contact model together with
the damping model developed by Tsuji et al. (1992) {[}\citealp{Tsuji1992}{]}
are used. In the tangential direction the force is described by a
spring limited by the Coulomb law characterised by the coefficient
of tangential friction {[}\citealp{Dziugys2001}{]}. In the present
study, only spherical particles are considered. A more detailed description
of the used DEM model can be found in {[}\citealp{Emden2014,Markauskas2015}{]}.

\subsection{Governing equations of the fluid phase\label{sec:Governing-eq-fluid}}

The Smoothed Particle Hydrodynamics (SPH) method {[}\citealp{Gingold1977},
\citealp{Lucy1977}{]} is used for modelling of the fluid and is coupled
with the DEM. The SPH treats the fluid in a completly mesh-free fashion
in terms of a set of sampling points (particles) {[}\citealp{Monaghan2005}{]}.
SPH particles represent a finite mass of the discretized fluid and
carry information about all physical variables evaluated at their
positions. Hydrodynamic equations for motion are derived for these
particles. 

The continuity equation and the momentum equation in a Lagrangian
framework take the form {[}\citealp{Robinson2014}{]}:

\begin{equation}
\frac{D\bar{\rho_{f}}}{Dt}+\nabla\cdot(\bar{\rho}_{f}\mathbf{u}_{f})=0,\label{eq:continuity-Lag}
\end{equation}

\begin{equation}
\frac{D\bar{\rho}_{f}\mathbf{u}_{f}}{Dt}=-\nabla p+\nabla\cdot(\varepsilon\boldsymbol{\tau})-\mathbf{f}^{int}+\bar{\rho}_{f}\mathbf{g}\textrm{,}\label{eq:momentum}
\end{equation}
where $\bar{\rho_{f}}=\varepsilon\rho_{f}$ is the superficial density
of the fluid, $\varepsilon$ is the local mean fluid volume fraction,
$\mathbf{u}_{f}$ is the fluid velocity, $p$ denotes the pressure,
$\boldsymbol{\tau}$ is the viscous stress tensor, $\mathbf{f}^{int}$
is the interaction force between fluid and solid particles and $\mathbf{g}$
is the gravitational constant.

SPH particles carry variables such as velocity, pressure and mass.
While no connectivity is modelled between the SPH particles, the function
values are interpolated from the neighbouring particles using a smoothing
kernel function. The kernel function is defined so that its value
monotonously decreases as the distance between particles increases.
The influence radius of the kernel function is defined by the smoothing
length $h$. There are several kernel functions used in SPH such as
the Gaussian {[}\citealp{Monaghan1983}{]}, quadratic {[}\citealp{Dalrymple2006}{]}
or quintic spline {[}\citealp{Wendland1995}{]}. In the current study
a commonly used cubic spline kernel {[}\citealp{Monaghan1999,Gomez2004}{]}
is utilised:

\begin{equation}
W(r,h)=\alpha_{D}\begin{cases}
1-\frac{3}{2}q^{2}+\frac{3}{4}q^{3}, & \quad0\leq q<1,\\
\frac{1}{4}(2-q)^{3}, & \quad1\leq q<2,\\
0, & \quad q\geq2,
\end{cases}\label{eq:KernelSpline}
\end{equation}
where $q=r/h$, $\alpha_{D}=1/(\pi h^{3})$ for the 3D case, $h$
is the smoothing length, which defines the influence volume of the
kernel and $r$ is the distance between the two points of interest.

An equation of state is used to estimate the pressure from the density
field in the weekly compressible SPH method {[}\citealp{Colagrossi2003,Monaghan2005}{]}:

\begin{equation}
p=\frac{\rho_{0}c^{2}}{\gamma}\left[\left(\frac{\bar{\rho_{f}}}{\varepsilon\rho_{0}}\right)^{\gamma}-1\right],\label{eq:state}
\end{equation}
where $\rho_{0}$ is the initial density of the fluid phase and $c$
is the speed of sound. It is recommended to use $c=10u$ to keep the
density to vary by at most 1\% {[}\citealp{Morris1997,Colagrossi2003}{]},
where $u$ is the maximum fluid velocity magnitude. The coefficient
$\gamma=7$ is commonly used in SPH. However we experienced numerical
instabilities when fluid together with solid particles were considered
in the simulations. Gao and Herbst {[}\citealp{Gao2009}{]} following
Morris et al. {[}\citealp{Morris1997}{]} recommend to use $\gamma=1$
to avoid these numerical problems. Following this recommendation $\gamma=1$
is used in the present study. 

The continuity equation (\ref{eq:continuity-Lag}) in the SPH takes
the form 

\begin{equation}
\frac{D\bar{\rho}_{a}}{Dt}=\underset{b}{\sum}m_{b}\mathbf{u}_{ab}\cdot\nabla_{a}W_{ab},\label{eq:continuity-SPH}
\end{equation}
where indexes $a$ and $b$ indicate fluid particles. $m$ is the
mass. $\mathbf{u}_{ab}=\mathbf{u}_{a}-\mathbf{u}_{b}$ is the relative
velocity between particles $\mathcal{P}_{a}$ and $\mathcal{P}_{b}$.
$\nabla_{a}W_{ab}=\nabla_{a}W(r_{a}-r_{b},h)$ is the gradient of
the kernel function. $r_{a}$ and $r_{b}$ are positions of the fluid
particles $\mathcal{P}_{a}$ and $\mathcal{P}_{b}$. The summation
is performed over all neighbouring particles of particle $\mathcal{P}_{a}$.

The momentum conservation equation (\ref{eq:momentum}) in SPH takes
the form {[}\citealp{Morris1997}{]}:

\begin{equation}
\begin{array}{c}
\frac{D\mathbf{u}_{a}}{Dt}=-\underset{b}{\sum}m_{b}\left(\frac{p_{a}}{\bar{\rho}_{a}^{2}}+\frac{p_{b}}{\bar{\rho}_{b}^{2}}\right)\nabla_{a}W_{ab}+\mathbf{g}+\\
+\underset{b}{\sum}m_{b}\frac{\nu(\bar{\rho}_{a}+\bar{\rho}_{b})}{\bar{\rho}_{a}\bar{\rho}_{b}}\cdot\frac{\mathbf{r}_{ab}\nabla_{a}W_{ab}}{|\mathbf{r}_{ab}|^{2}+\delta^{2}}\mathbf{u}_{ab}+\frac{\mathbf{f}_{a}^{int}}{m_{a}}.
\end{array}\label{eq:momentum-SPH}
\end{equation}

The third term on the right hand side in Eq. (\ref{eq:momentum-SPH})
is a viscous term introduced by Morris {[}\citealp{Morris1997}{]},
where $\nu$ is the kinematic viscosity. $\delta$ is a small number
used just to keep the denominator non-zero which here is set to $0.1h$.

$\mathbf{f}_{a}^{int}$ in Eq. (\ref{eq:momentum-SPH}) is the solid-fluid
interaction force acting on the fluid particle $\mathcal{P}_{a}$
due to the solid particles. The force $\mathbf{f}_{a}^{int}$ is calculated
as the sum over all solid particles in the domain of the fluid particle:

\begin{equation}
\mathbf{f}_{a}^{int}=\underset{i}{\sum}-\frac{V_{a}W_{ai}}{\underset{b}{\sum}V_{b}W_{bi}}\mathbf{F}_{i}^{int}\:.\label{eq:f_int_S-F}
\end{equation}
where $V_{a}$ is the volume of fluid particle, while $\mathbf{F}_{i}^{int}$
is the interaction force acting on the solid particle (see Eq. (\ref{eq:sp_newton})).

The fluid volume fraction $\varepsilon_{a}$ of the fluid particle
$\mathcal{P}_{a}$ is calculated from the volumes of all solid particles
$\mathcal{P}_{i}$ which are in the smoothing domain of the fluid
particle $\mathcal{P}_{a}$:

\begin{equation}
\varepsilon_{a}=1-\sum_{i}V_{i}W_{ai}\:,\label{eq:fluid-fraction}
\end{equation}
where $V_{i}$ is the volume of the solid particle $\mathcal{P}_{i}$,
while $W_{ai}=W(r_{a}-r_{i},h)$ is the kernel function Eq. (\ref{eq:KernelSpline}).

Fluid particles are moved using a velocity smoothed by the average
in their neighbourhood according to the kernel function, i.e. XSPH
variant introduced by Monaghan {[}Mon1989{]}:

\[
\frac{d\mathbf{r}}{dt}=\mathbf{v}_{a}+\varepsilon_{XSPH}\sum\frac{m_{b}}{\hat{\rho}_{ab}}(\mathbf{v}_{b}-\mathbf{v}_{a})W_{ab},
\]
where $\varepsilon_{XSPH}$ is the parameter here used equal to 0.5
and $\hat{\rho}_{ab}=(\rho_{a}+\rho_{b})/2$. This smoothed particle
velocity reduces the fluid particle disorder, while does not change
the overall linear momentum.

An often discussed topic in the SPH literature is the use of boundary
conditions {[}\citealp{Adami2012,Marrone2013,Valizadeh2015}{]}. It
is related to the fact, that the description of boundaries in the
SPH is not as straightforward as in other grid based CFD methods.
There are several approaches to enforce no-penetration boundaries
in the SPH, in which mostly special fluid wall particles are introduced.
In our earlier study {[}\citealp{Markauskas2017}{]} a modification
of a no-slip no-penetration boundary was proposed, in which instantaneously
generated ghost-fluid particles were used. These boundaries performed
well in the test cases in {[}\citealp{Markauskas2017}{]}. However
in the current study, where particle separation in rotating drum is
simulated, we experienced numerical problems when the fluid particles
are moved above the free surface of the fluid by the lifters (Fig.\,\ref{fig:Scheme-Drum}).
In this situation a negative pressure (tension) in some of the fluid
particles arose, which caused an artificial attraction between those
particles and the nearby walls. To prevent this problem, a repulsive
force boundary model proposed by Monaghan {[}\citealp{Monaghan1994}{]}
is used in the current study. For this boundary condition the force
between the fluid particle and the wall does not depend on the fluid
particle pressure. Therefore, no artificial tension between a particle
and a wall is generated. The same type of boundary condition was used
by Gao et al. {[}\citealp{Gao2009}{]} and Robinson et al. {[}\citealp{Robinson2014}{]}
in their studies applying the coupled DEM-SPH method. 

\subsection{Fluid-solid interaction\label{sec:Interaction}}

The interaction force acting on a solid particle $\mathbf{F}_{i}^{int}$
in this study is calculated as the sum of the drag force $\mathbf{F}_{i}^{D}$
and the pressure gradient force $\mathbf{F}_{i}^{\nabla p}$:

\begin{equation}
\mathbf{F}_{i}^{int}=\mathbf{F}_{i}^{D}+\mathbf{F}_{i}^{\nabla p}.\label{eq:Interaction}
\end{equation}

In the current study the correlation proposed by Di\,\,Felice {[}\citealp{DiFelice1994}{]},
which is well-anticipated in literature, is used for the calculation
of the drag force:

\begin{equation}
\mathbf{F}_{i}^{D}=\frac{1}{8}C_{d}\rho_{f}\pi d_{i}^{2}(\mathbf{u}_{f,i}-\mathbf{v}_{i})|\mathbf{u}_{f,i}-\mathbf{v}_{i}|\varepsilon_{i}^{2-\chi},\label{eq:DiFelice}
\end{equation}
where $\varepsilon_{i}$, $d_{i}$, $\mathbf{u}_{f,i}$, $\mathbf{v}_{i}$
are the fluid fraction at the location of solid particle $\mathcal{P}_{i}$,
the solid particle diameter, the fluid velocity and the solid particle
velocity correspondingly. $\varepsilon_{i}$ is obtained from the
fluid fractions at the surrounding fluid particles:

\begin{equation}
\varepsilon_{i}=\frac{\underset{a}{\sum}\varepsilon_{a}V_{a}W_{ai}}{\underset{a}{\sum}V_{a}W_{ai}}\thinspace.
\end{equation}
The drag coefficient $C_{d}$ and the coefficient $\chi$ are calculated
as a function of the particle Reynolds number {[}\citealp{DiFelice1994}{]}. 

Assuming that the pressure gradient $\nabla p$ arises only because
of the interaction between solid particles and fluid, $F_{i}^{D}$
can be combined with $F_{i}^{\nabla p}$ {[}\citealp{Oschmann2014}{]},
which results in:

\begin{equation}
\mathbf{F}_{i}^{int}=\frac{\mathbf{F}_{i}^{D}}{\varepsilon}-V_{i}\rho_{f}\mathbf{g}.
\end{equation}

$\mathbf{F}_{i}^{int}$ is used in Eq. (\ref{eq:sp_newton}) and Eq.
(\ref{eq:f_int_S-F}). More details about the used DEM-SPH model can
be found in {[}\citealp{Markauskas2017}{]}.

\section{Verification of the numerical model \label{sec:Testing-NumMod}}

For verification purposes of the theoretical model presented in section
\ref{sec:Governing-eq}, numerical tests of a dam break problem are
performed. In subsection \ref{subsec:DamB-Single} a single phase
(liquid only) dam break problem, while in subsection \ref{subsec:DamB-Two}
a two phase (liquid and solid particles) dam break problem are simulated
and obtained results are compared with available results found in
literature.

\subsection{Dam break: Single-phase test \label{subsec:DamB-Single}}

\begin{figure}
a)\includegraphics[width=7.5cm]{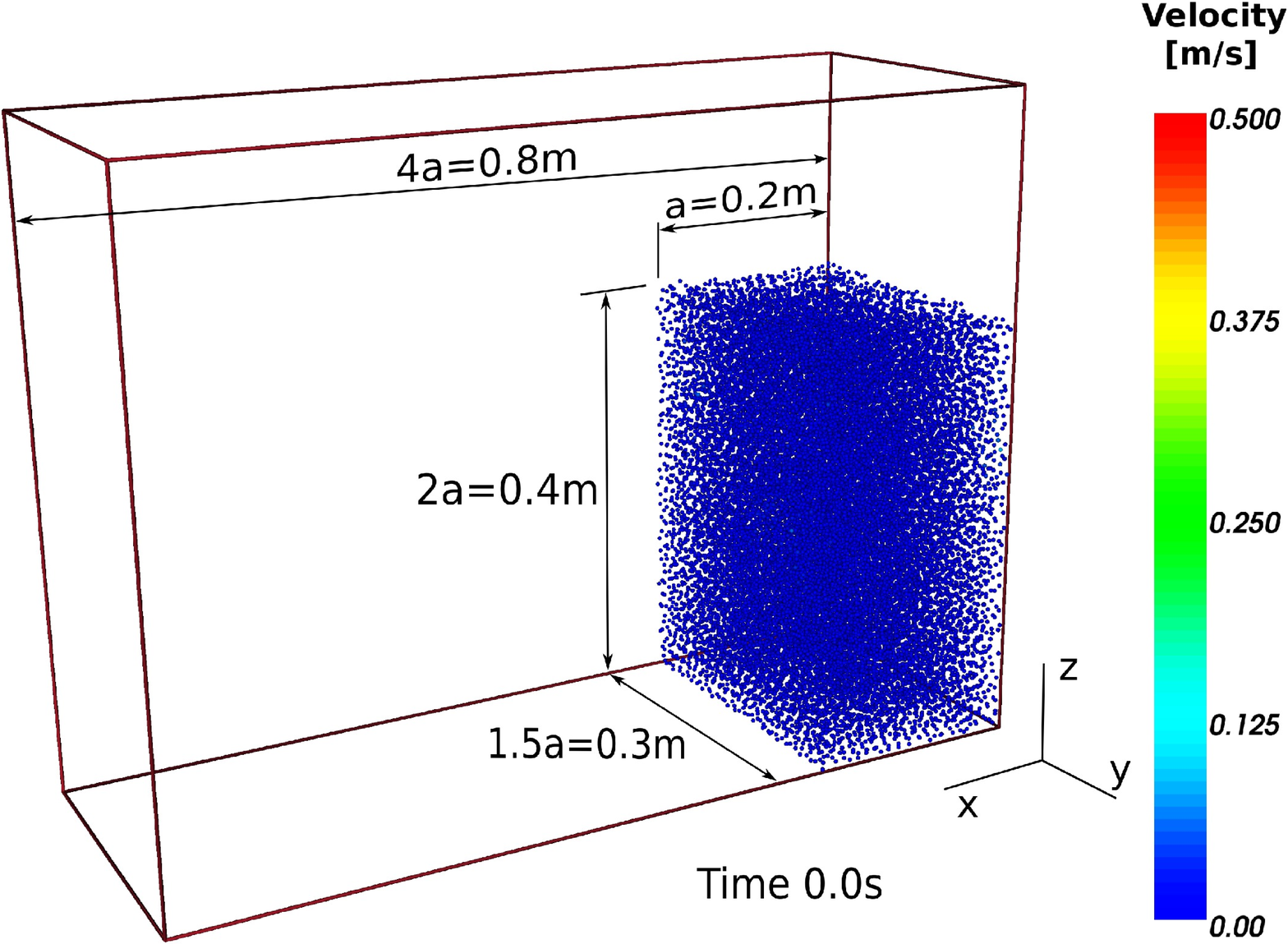}
b)\includegraphics[width=7.5cm]{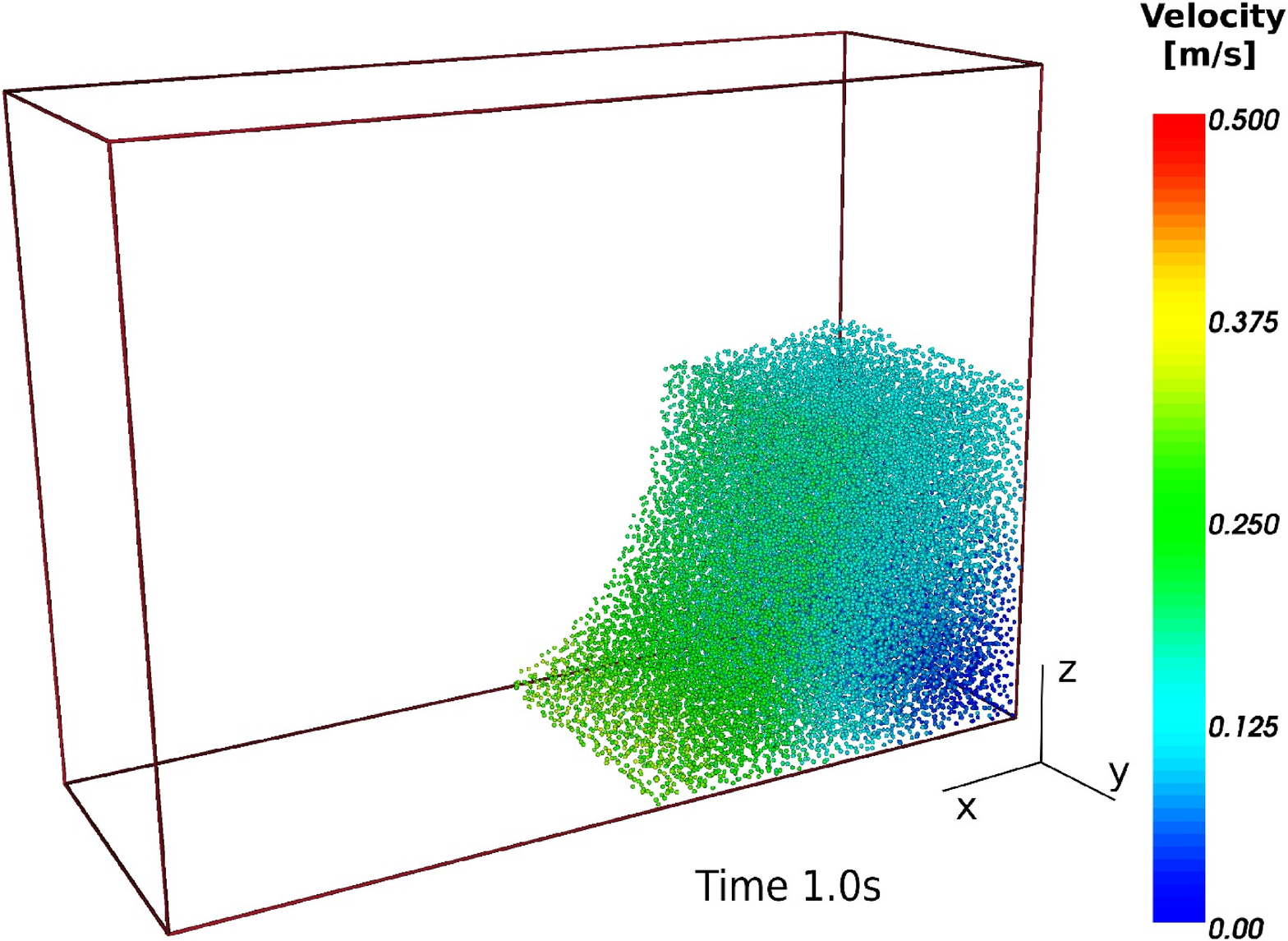}
c)\includegraphics[width=7.5cm]{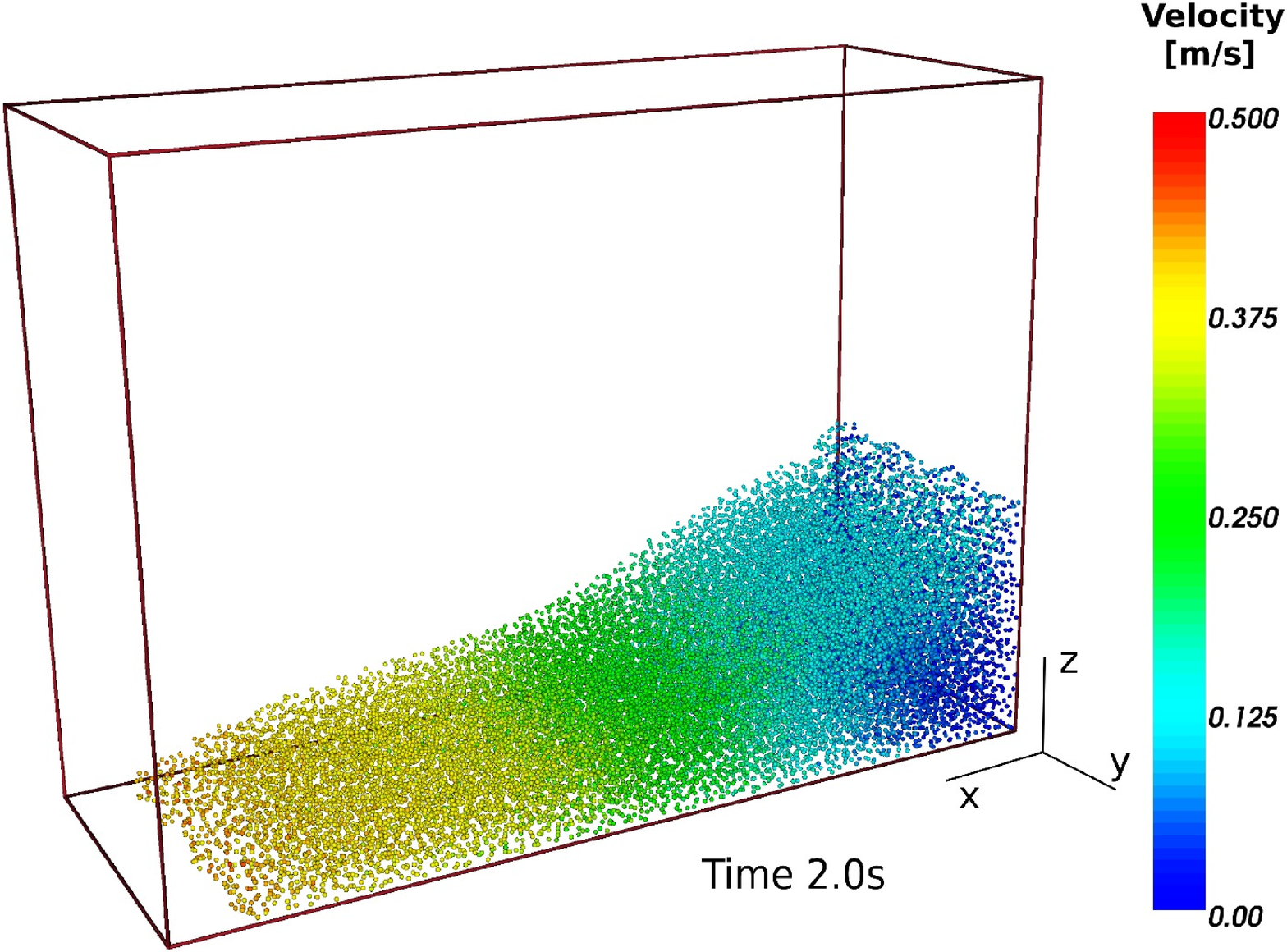}
d)\includegraphics[width=7.5cm]{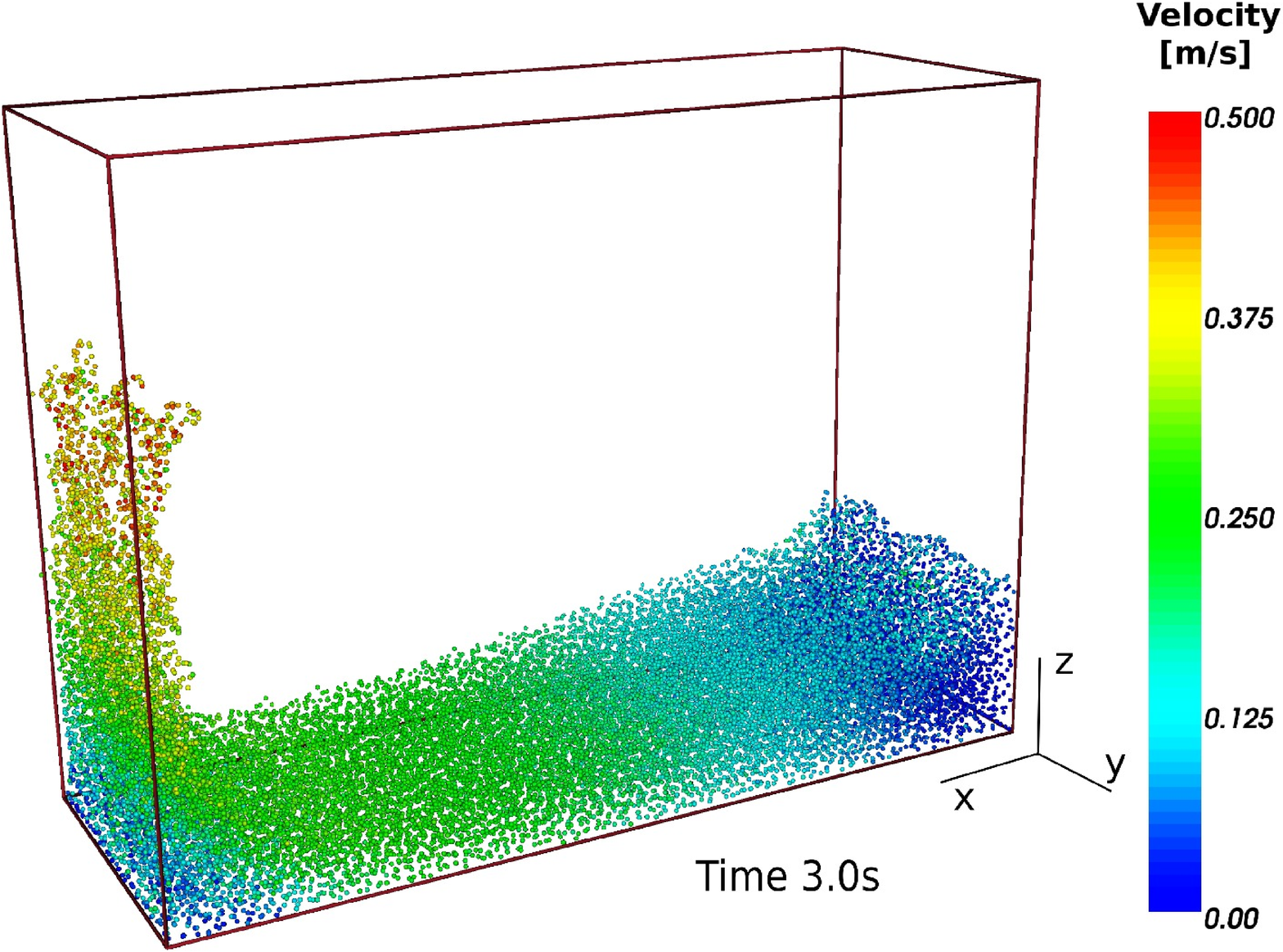}

\caption{SPH particles representing the liquid in the single-phase dam break
test\label{fig:DamBSin_Part}}

\end{figure}

\begin{figure}
\begin{centering}
\includegraphics[width=7.5cm]{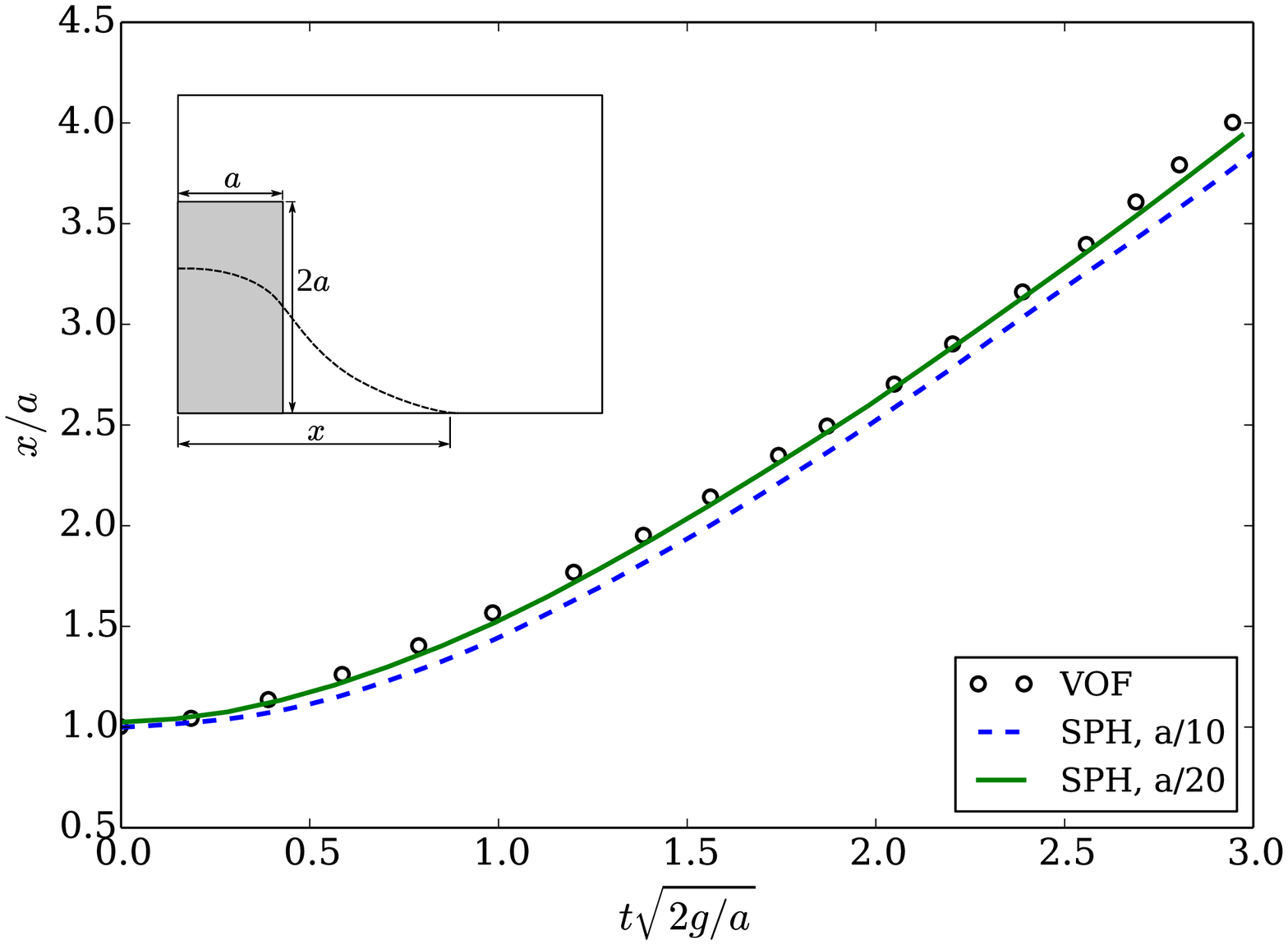}
\par\end{centering}
\caption{Position of the leading edge of the liquid in the single-phase dam
break test \label{fig:DamBSin_edge}}

\end{figure}

In this test, a part of a rectangular container is filled with water
(Fig. \ref{fig:DamBSin_Part}). The width of the liquid column is
$a=0.2\thinspace\mathrm{m}$ and the height is $2a=0.4\thinspace\mathrm{m}$.
The gravity force is acting downwards with the magnitude $a$, i.e.
$0.2\thinspace\mathrm{m/s^{2}}$. At the start of the simulation,
an initialisation step is performed, during which the liquid particles
are allowed to reach a static equilibrium condition. Then the right
wall of the container is removed and the liquid is flowing along the
horizontal bottom plane as a consequence. The 2D numerical test thereby
using the volume of fluid (VOF) method was initially performed and
reported by Hirt and Nichols {[}\citealp{Hirt1981}{]}. The same test
was also used by Sun et al. {[}\citealp{Sun2013}{]}. Two simulations
are performed: one using an initial distance between liquid particles
equal to $a/10$ and one using an initial distance between liquid
particles equal to $a/20$. Totally 3000 and 24000 SPH particles are
used for the first and for the second simulation respectively. Liquid
particle positions for the $a/20$ simulation at different time instances
are shown in Fig.\,\ref{fig:DamBSin_Part}. The position of the leading
edge of the liquid vs time is presented in Fig.\,\ref{fig:DamBSin_edge},
where SPH results performed with two different SPH particle resolutions
together with VOF results from {[}\citealp{Hirt1981}{]} can be compared. 

It can be seen in Fig.\,\ref{fig:DamBSin_edge}, that when using
$a/10$ SPH particles, the difference between VOF and SPH results
reaches 5\,\% at $t\sqrt{2g/a}=2.0$. The reduction of the size of
the SPH particles by a factor of two ($a/20$) reduces this difference
down to 1\,\%. The obtained results let us conclude, that the used
SPH model reproduces VOF results very well.

\subsection{Dam break: Two-phase test\label{subsec:DamB-Two}}

\begin{figure}
\begin{centering}
a)\includegraphics[width=7cm]{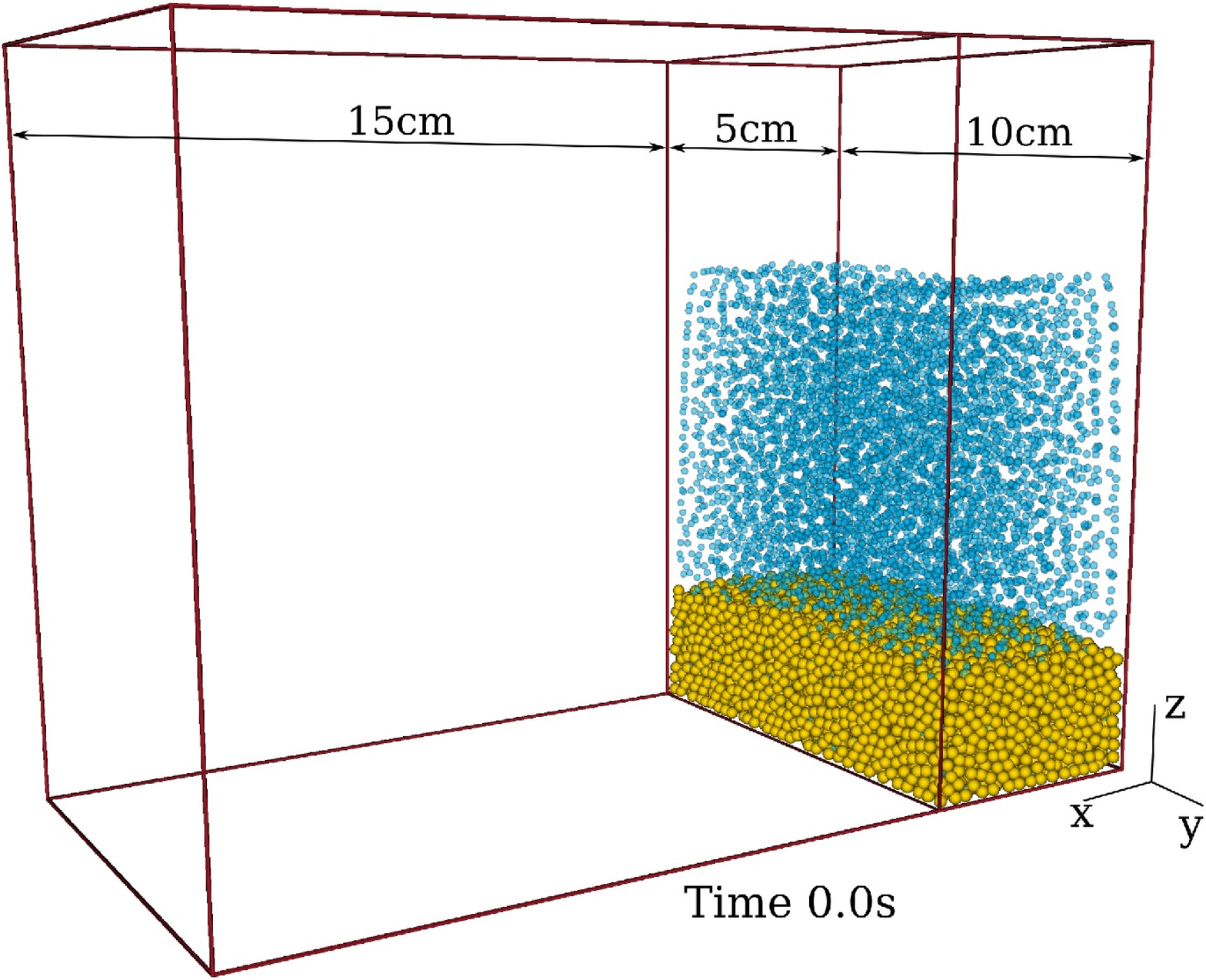} b)\includegraphics[width=7cm]{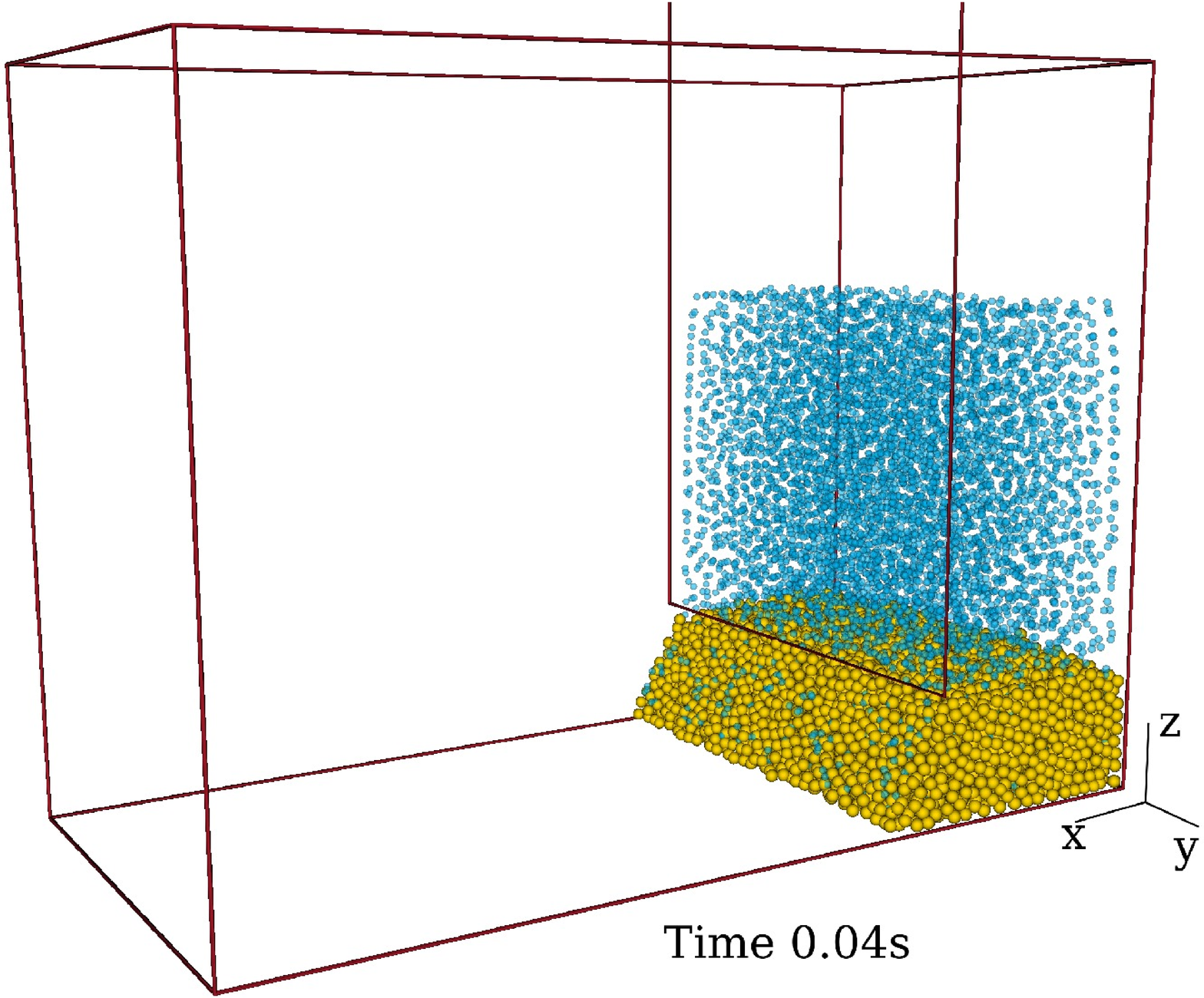}
c)\includegraphics[width=7cm]{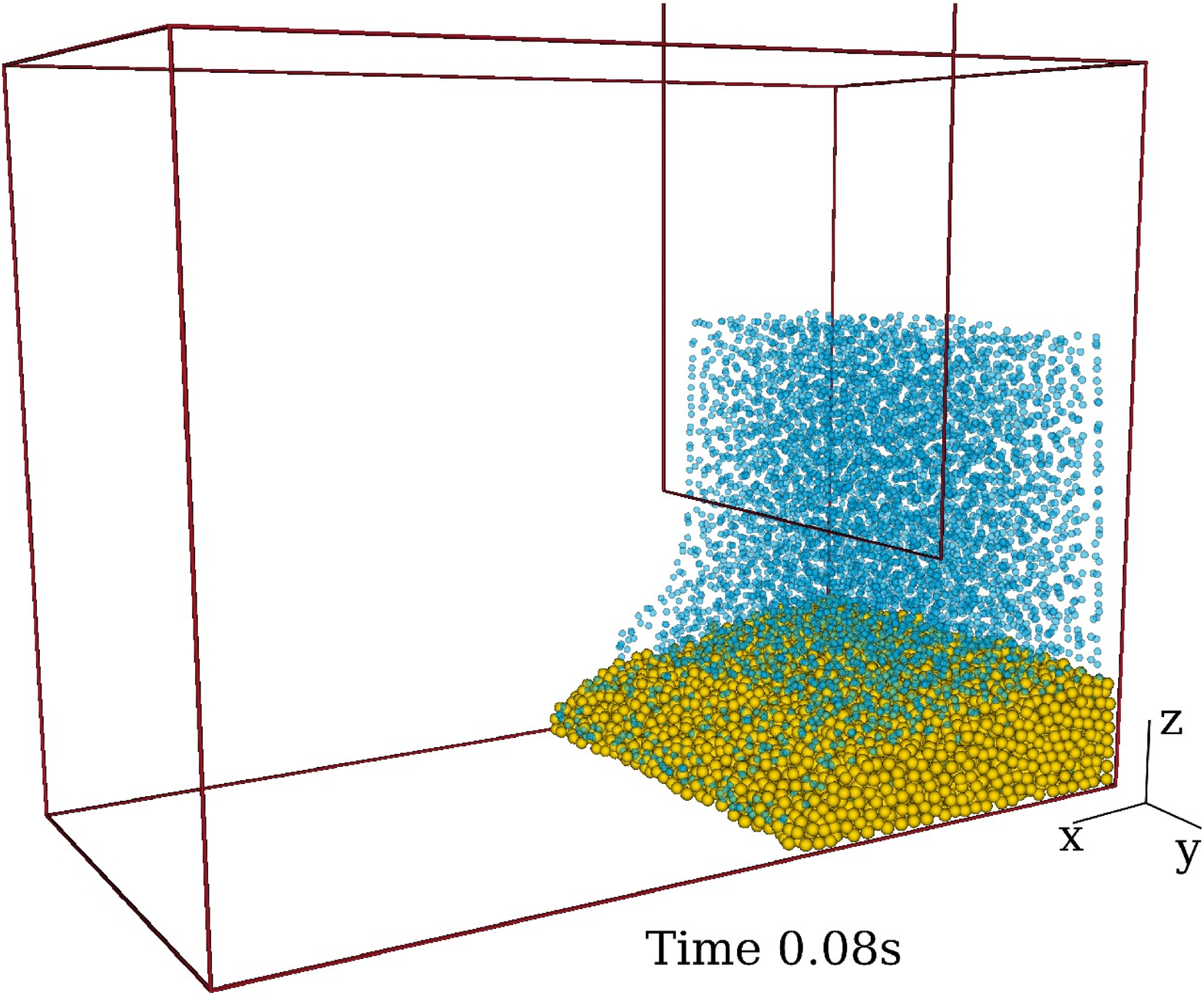} d)\includegraphics[width=7cm]{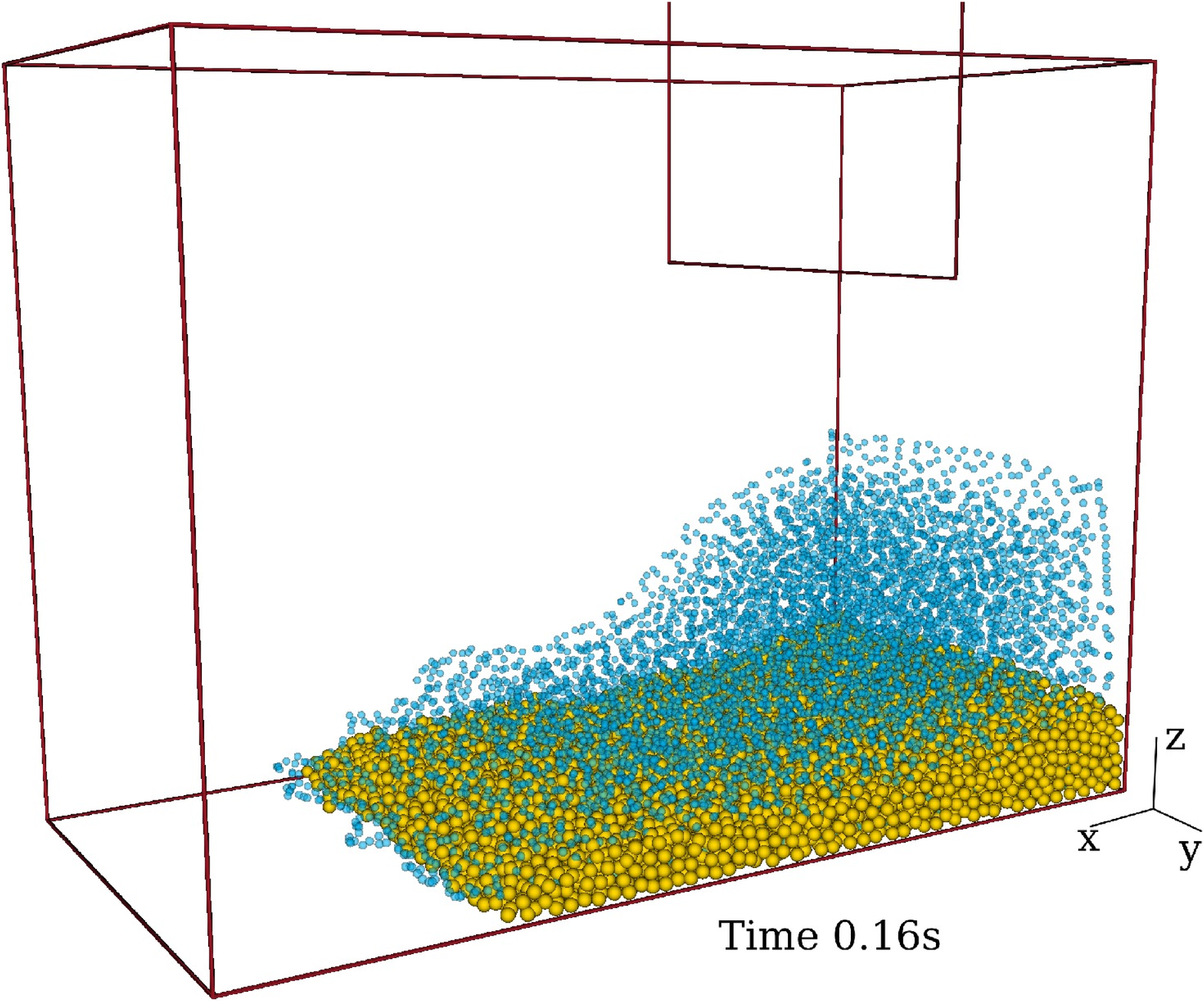}
\par\end{centering}
\caption{SPH particles (representing the liquid) and solid particles in the
two-phase dam break test\label{fig:DamBTwo_Part}}
\end{figure}

\begin{figure}
\begin{centering}
\includegraphics[width=7.5cm]{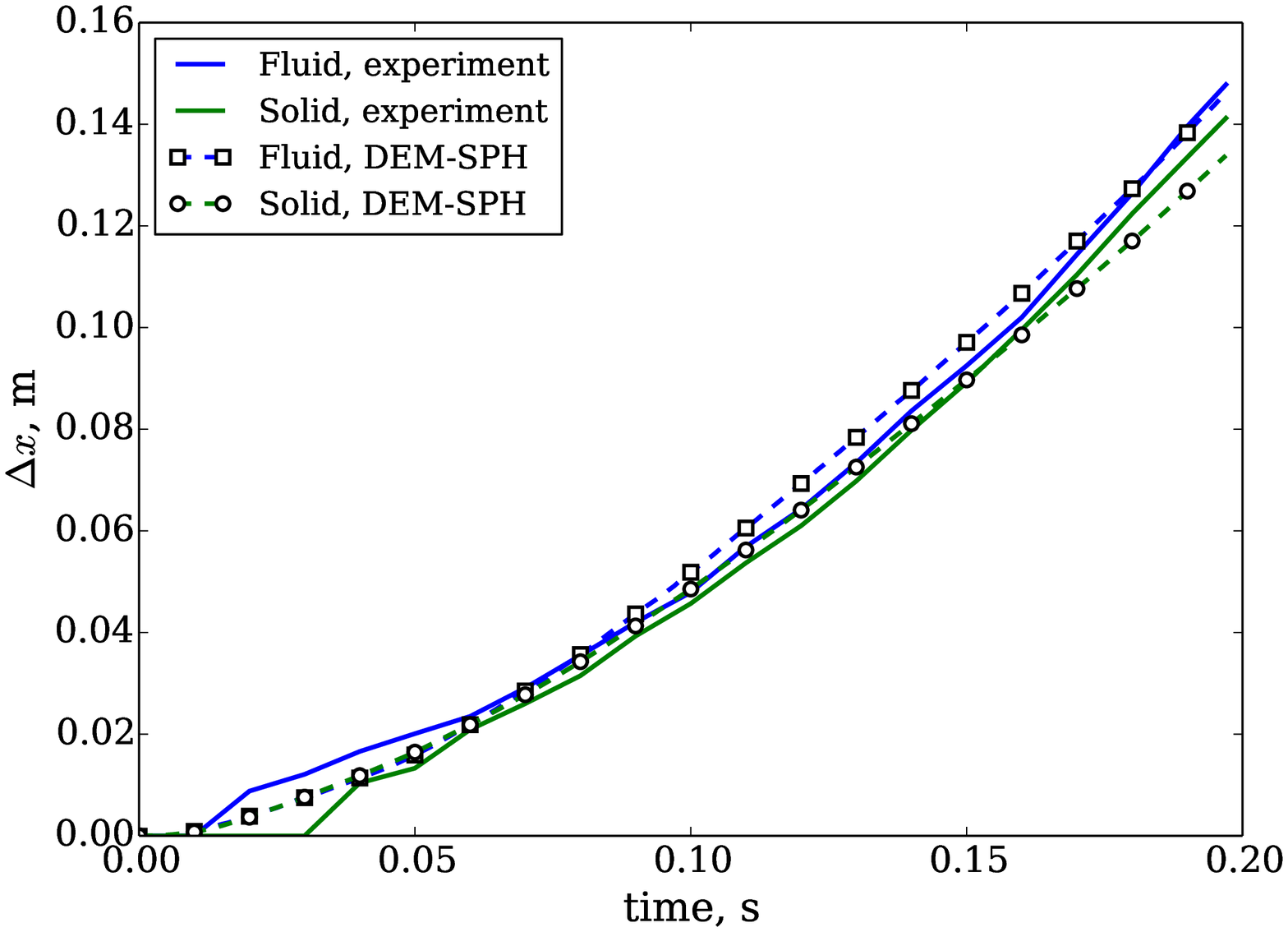}
\par\end{centering}
\caption{Temporal variation of the position of the leading edge of the the
fluid and solid particles in the two-phase dam break test \label{fig:DamBTwo_edge}}

\end{figure}

A simulation of the dam break problem, in which water is laden with
solid particles, is performed as a second test case. The obtained
numerical results are compared with experimental results reported
by Sun et al. {[}\citealp{Sun2013}{]}. A rectangular container with
size $20\,\mathrm{cm}\times10\,\mathrm{cm}\times15\,\mathrm{cm}$
(see Fig.\,\ref{fig:DamBTwo_Part}) is divided into two sections
(one larger and one smaller section) by a vertical wall. In a first
step, solid particles are generated with fluid particles placed atop
of them in the smaller section of the container (Fig. \ref{fig:DamBTwo_Part}\,a).
Spherical solid particles with diameter 2.7\,mm and density $2500\,\mathrm{kg/m^{3}}$
are used for modelling of the glass beads. 200\,g of solid material,
which results in 7762 particles, is generated. The particle's Young's
modulus equals 100\,MPa, its Poisson's ratio is 0.2, its restitution
coefficient is set to 0.9 and the friction coefficient equals 0.2
as reported for the solid particles in the experiment {[}\citealp{Sun2013}{]}.
For modelling of the water phase, which is used in {[}\citealp{Sun2013}{]}
in the performed experiment, 5870 SPH particles with a density of
$\rho_{0}=1000\,\mathrm{kg/m^{3}}$, a dynamic viscosity $\mu=0.001\,\mathrm{Pa\cdot s}$,
a smoothing radius $h=5.4\,\mathrm{mm}$ and an initial distance between
SPH particles equal to $h/1.3$ are used. An initialisation is performed
during which the solid particles and the liquid are allowed to settle
down by the action of gravity. In the actual simulation, the vertical
wall (dam), which divides the container into two sections, is raised
by a constant velocity of $v_{x}=0.68\,\mathrm{m/s}$, therefore the
solid particles-fluid mixture is moving out from the filled section.
The change of the position of the leading edge of the fluid and the
solid particles is shown in Fig.\,\ref{fig:DamBTwo_edge}.

As can be seen in Fig.\,\ref{fig:DamBTwo_edge}, the simulation results
match well with the experiment. Some minor divergence of the positions
of the solid particles obtained numerically from the reported experimentally
measured values at $t=0.16-0.20\,\mathrm{s}$ can be observed from
the presented curves, however. The possible reason for this minor
divergence could be in the physical properties of solid particles
(restitution coefficient, friction coefficient) which where just roughly
estimated in Ref.\,{[}\citealp{Sun2013}{]}. 

\section{Numerical analysis of wet plastic particle separation \label{sec:NumAnal-separation}}

\subsection{Simulation setup and parameters\label{s:Simul_Basic}}

A numerical analysis of the separation of plastic particles using
a rotating drum is performed. The outline of the drum is presented
in Fig.\,\ref{fig:Scheme-Drum}. During the separation process the
mixture of grains together with water is fed into the rotating drum
through the opening on the right side (Fig.\,\ref{fig:Scheme-Drum}
a). By interaction of gravity and buoyancy forces, the grains with
a density lower than the liquid density are floating, while the grains
with a density higher than the liquid density start to sink. The floating
grains together with the liquid are discharged out through the opening
on the left side of the drum. The sunken particles are lifted by lifters
attached to the walls of the drum and dropped on the sink launder.

The separation of polyethylene terephthalate (PET) from polypropylene
(PP) is simulated. The density of PET particles is considered with
$1350\,\mathrm{kg/m^{3}}$ and the density of PP particles with $950\,\mathrm{kg/m^{3}}$;
a restitution coefficient of 0.5 is used. A density of $1000\,\mathrm{kg/m^{3}}$
and a dynamic viscosity of $0.001\,\mathrm{Pa\cdot s}$ are used for
the liquid aligned with the properties of water.

\begin{figure}
\begin{centering}
\includegraphics[width=7.5cm]{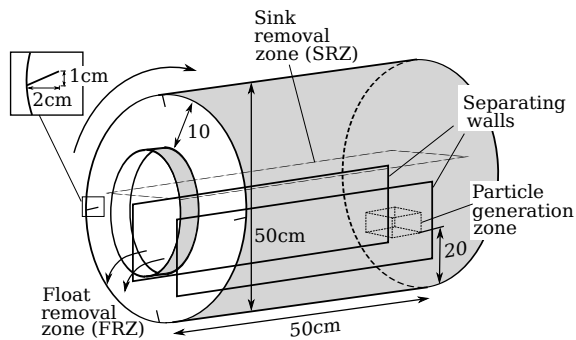}
\par\end{centering}
\caption{Scheme of the drum separator used in the simulations \label{fig:DrumSimple}}
\end{figure}

In the simulations a simplified scheme of the laboratory scale drum
shown in Fig.\,\ref{fig:DrumSimple} is used. Initially a prefill
of the drum is numerically performed, where $0.0198\,\mathrm{m^{3}}$
of water are generated inside of the drum and a simulation of 1\,s
is performed during which this water settles and partly flows out
of the drum forming a starting condition. Then the mixture of solid
particles and fluid is started to be generated inside the drum in
small chunks every 0.09\,s. By varying the size of the chunk a possible
variation of the flow rate is achieved. In all performed simulations
the particle mixture contains an equal volume of PET and PP material.
For the modelling of the PET and PP material, spherical particles
with random diameter between 3-4\,mm are used. For the modelling
of the water, SPH particles with a kernel length of $h=8\,\mathrm{mm}$
and an initial distance of $h/1.3$ are utilised. The kernel length
is chosen based on the simulation results described in \citep{Robinson2014}
and the analysis of the settlement of a single particle in \citep{Markauskas2017}.
All simulations are performed until 40\,s of operation time is reached.
During the simulation the sinking/sunken solid particles are lifted
from the bottom of the drum and are dropped into the ``Sink Remove
Zone'' (SRZ) (see Fig.\,\ref{fig:DrumSimple}), where particles
are removed from the simulation. Thereby the discharge process of
the sinking/sunken solid particles is represented in a simplified
manner which however has no implication on the accuracy of the simulation.
The floating solid particles together with the water flow out through
the opening in the drum and are also removed (``Float Remove Zone''
(FRZ) in Fig.\,\ref{fig:DrumSimple}) . 

Overall 10 simulations are performed with the purpose to analyse the
sensitivity of the separation quality to different parameters. The
considered parameters are given in Table\,\ref{tab:SensParam}. As
a base setup (BS in Table\,\ref{tab:SensParam}) a simulation with
a rotational velocity of the drum of $0.5\pi\,\mathrm{rad/s}$, 4
lifters, a solid feed rate of 29.3\,g/s and a water flow rate of
$2.07\cdot10^{-3}\,\mathrm{m^{3}/s}$ is utilised. Also vertical walls
inside of the drum (see Fig.\,\ref{fig:DrumSimple}) are used to
form a zone where the separation of solid particles should take place.
Every simulation S1-S9 differs from the BS just by one parameter with
a two times smaller and two times larger value being applied. In Table\,\ref{tab:SensParam}
only the parameters for the BS and the parameters for S1-S9, which
differ from the BS, are presented.

\begin{table}
\caption{Simulation parameters for testing of the drum separation \label{tab:SensParam}}

\begin{tabular*}{17.5cm}{@{\extracolsep{\fill}}>{\centering}p{2.5cm}>{\centering}p{2.5cm}>{\centering}p{2.5cm}>{\centering}p{2.5cm}>{\centering}p{2.5cm}>{\centering}p{2.5cm}}
\hline 
Label & Separating walls & Rotational velocity of the drum, rad/s & Number of lifters & Solid feed rate, g/s & Water flow rate, $\mathrm{m^{3}/s}$\tabularnewline
\hline 
BS & Yes & 0.5 & 4 & 29.3 & $2.07\cdot10^{-3}$\tabularnewline
S1 & No &  &  &  & \tabularnewline
S2 &  & $0.25\pi$ &  &  & \tabularnewline
S3 &  & $\pi$ &  &  & \tabularnewline
S4 &  &  & 2 &  & \tabularnewline
S5 &  &  & 8 &  & \tabularnewline
S6 &  &  &  & 14.65 & \tabularnewline
S7 &  &  &  & 58.6 & \tabularnewline
S8 &  &  &  &  & $1.035\cdot10^{-3}$\tabularnewline
S9 &  &  &  &  & $4.14\cdot10^{-3}$\tabularnewline
\hline 
\end{tabular*}
\end{table}

\subsection{Simulation of basic setup}

\begin{figure}
\begin{centering}
a)\includegraphics[width=6.5cm]{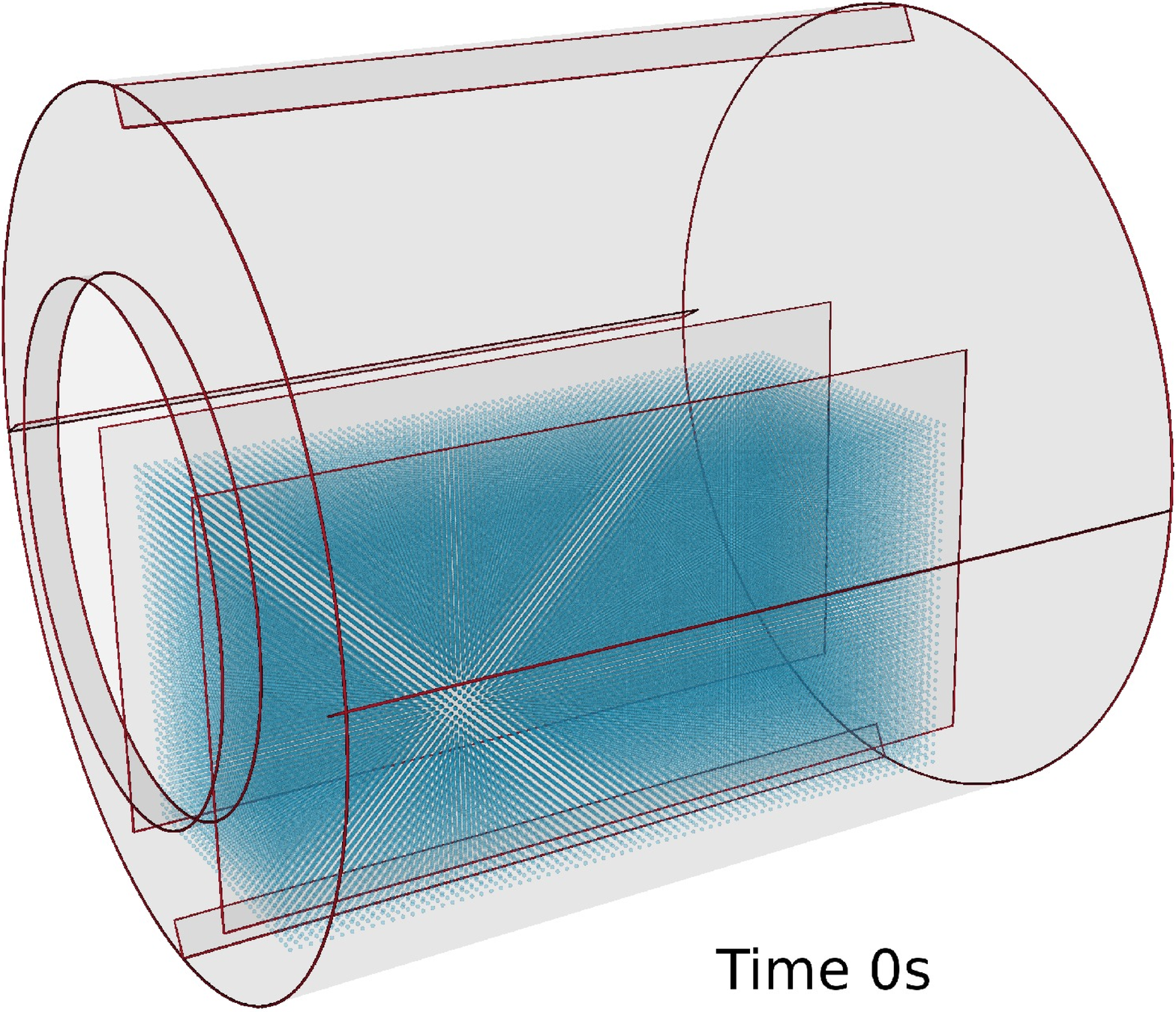} b)\includegraphics[width=6.5cm]{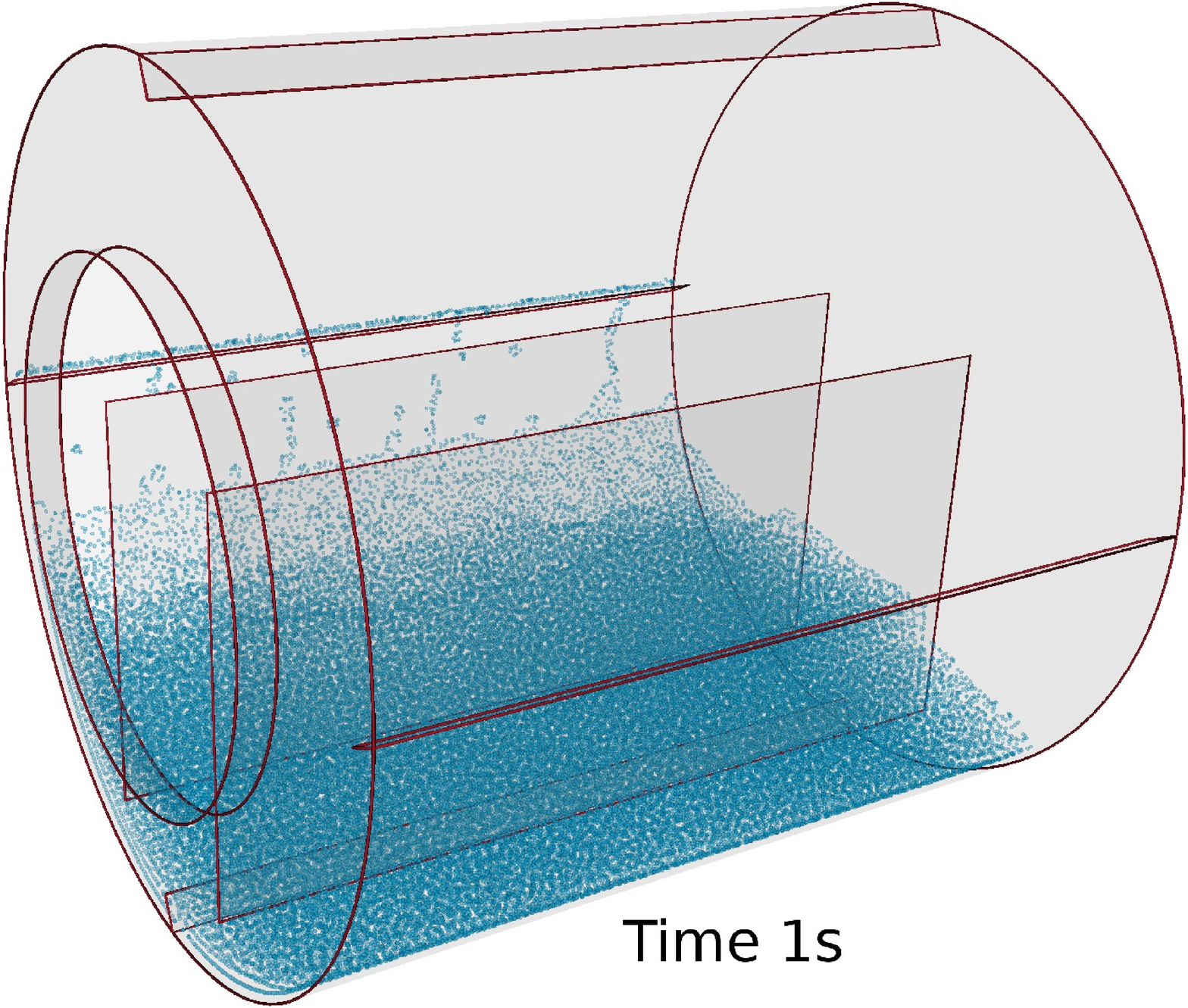}
c)\includegraphics[width=6.5cm]{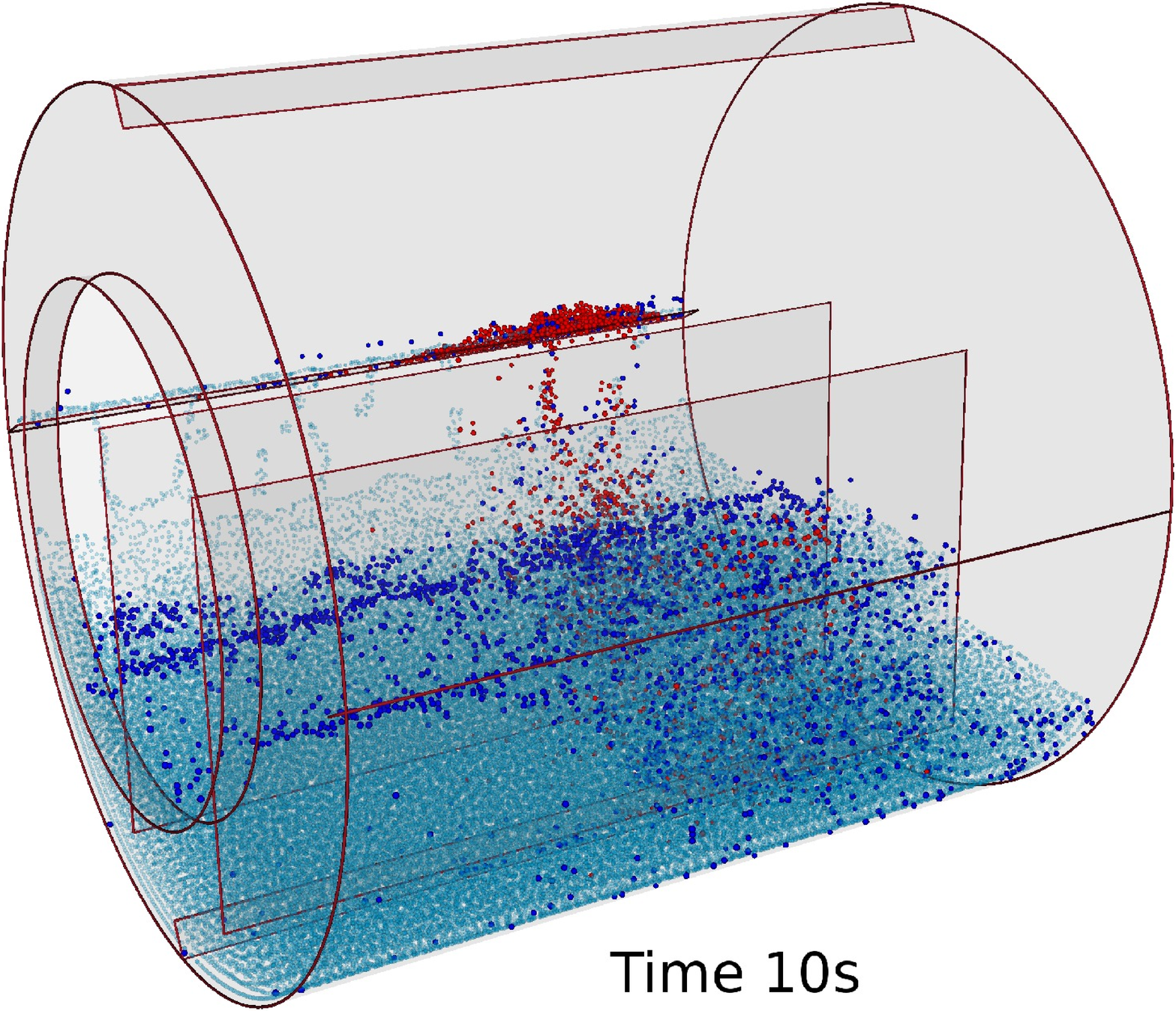} d)\includegraphics[width=6.5cm]{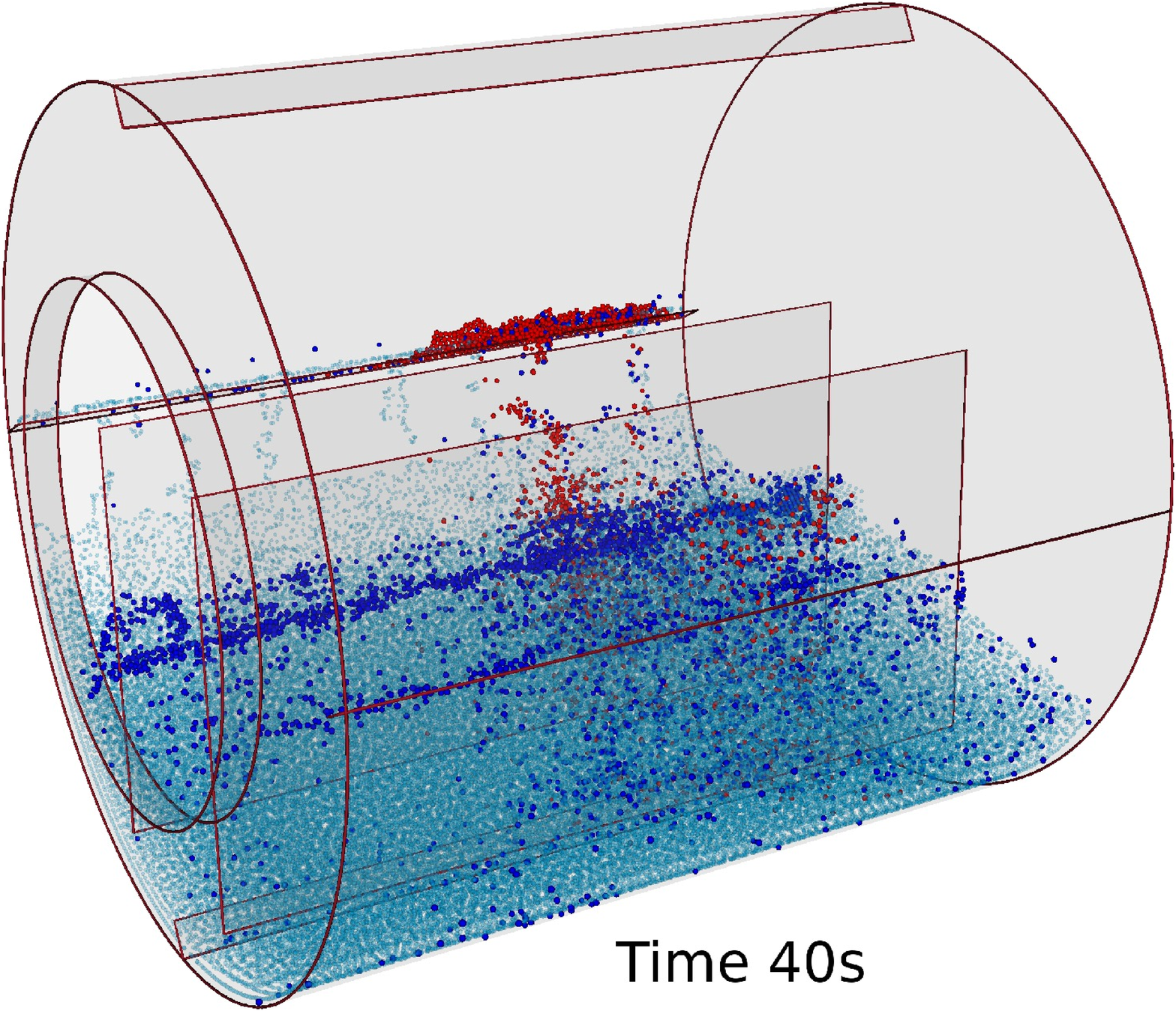}
\par\end{centering}
\caption{Simulation of particle separation in the rotating drum: snapshots
taken at different time instances \label{fig:Drum_SB}}

\end{figure}

Several snapshots of the simulation of the basic setup (BS in Table\,\ref{tab:SensParam})
at different instances of time are presented in Fig.\,\ref{fig:Drum_SB}.
In the figures, the SPH water particles are shown in a light blue
colour and are semi transparent. PET particles ($\rho=1350\,\mathrm{kg/m^{3}}$)
are shown in a red colour and PP particles ($\rho=950\,\mathrm{kg/m^{3}}$)
are represented dark blue. As already mentioned in Section \ref{s:Simul_Basic},
the drum is initially filled with water. The column of water is generated
inside of the drum at $t=0\thinspace\mathrm{s}$ (Fig.\,\ref{fig:Drum_SB}\,a),
which is allowed to settle down until 1\,s (Fig.\,\ref{fig:Drum_SB}\,b).
Then the mixture of water and solid particles is started to be generated
inside the drum. In Fig.\,\ref{fig:Drum_SB}\,c and Fig.\,\ref{fig:Drum_SB}\,d
it can be seen how solid particles are moved up out of the liquid
by the lifters. However not only sunken PET particles (red colour),
but some water and floating PP particles (blue colour) are caught
and transported to the Sink Remove Zone (SRZ) (see Fig.\,\ref{fig:DrumSimple}).
Here the fluid and liquid particles are removed from the simulation,
which has no effect on the separation process as even in reality they
would leave the system here.

\begin{figure}
\begin{centering}
\includegraphics[width=7.5cm]{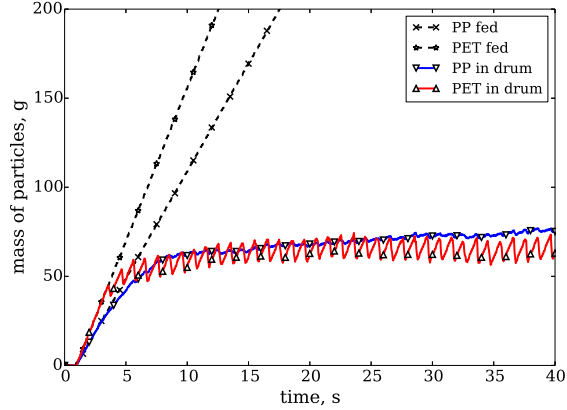}
\par\end{centering}
\caption{Basic setup (BS) simulation: the mass of the generated particles and
the mass of the particles inside of the drum \label{fig:SB_gen_inside}}

\end{figure}

The mass of the fed plastic particles and the mass of the plastic
particles inside the drum during the simulation are shown in Fig.\,\ref{fig:SB_gen_inside}.
While in the figure the shown mass of the fed particles is cut at
200\,g, the particles are charged constantly until the end of the
simulation. The mass of PET particles inside of the drum at first
increases rapidly, however from about $t=6\,\mathrm{s}$, the amount
of PET particles inside remains almost unchanged. The stepwise character
of the ``PET in drum'' curve reflects the time intervals at which
the rotating lifters remove sunken PET particles. The amount of PP
particles increases rapidly until about $t=8\,\mathrm{s}$, but then
a light steady increase is forming out which remains until the end
of the simulation.

\begin{figure}
\begin{centering}
\includegraphics[width=7.5cm]{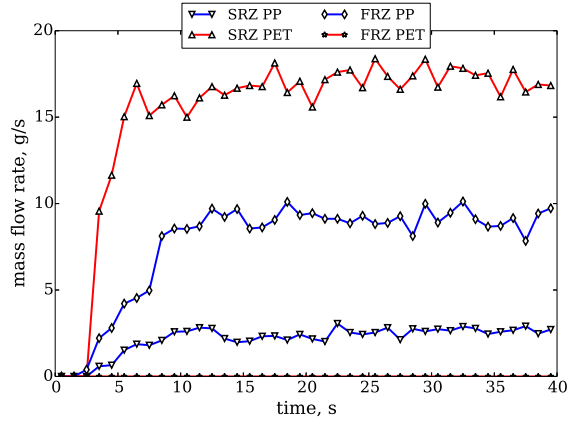}
\par\end{centering}
\caption{Mass flow rate of the separated particles in the basic simulation
(BS) \label{fig:SB_Mass-rate}}
\end{figure}

From the amount of the removed particles a mass flow rate is calculated.
Because particles can be removed either when they are lifted to the
sink remove zone (SRZ), or when they float through the left opening
in the drum (float remove zone, FRZ) (see Fig.\,\ref{fig:DrumSimple}),
two mass rates for every particle kind are obtained and shown in Fig.\,\ref{fig:SB_Mass-rate}.
In an ideal separation case, all PET particles should reach the SRZ,
while all PP particles should exit through the FRZ. While no PET particle
is transported to the FRZ, in the basic simulation case (BS) some
PP particles are moving into the SRZ together with PET. Therefore,
the resulting ``SRZ PP'' curve is non-zero.

\subsection{Analysis of the influence of different design and operational parameters}

\subsubsection{Influence of separating walls}

\begin{figure}
\begin{centering}
\includegraphics[width=7.5cm]{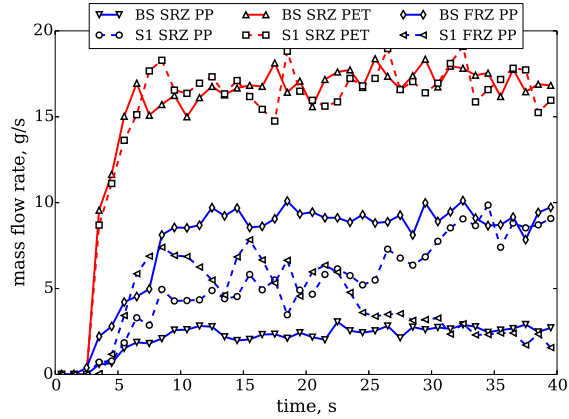}
\par\end{centering}
\caption{The mass flow rate of particle removal for S1 and BS simulations \label{fig:Mass-rate-S2}}

\end{figure}

The influence of separating vertical walls (see Fig.\,\ref{fig:DrumSimple})
is analysed. In many drum separators they are used to confine the
zone of the settling of particles. Besides the basic setup (BS) simulation,
in which vertical walls are used, a simulation without vertical walls
(S1 in Table\,\ref{tab:SensParam}) is performed. The development
of the mass flow rate of particle removal is shown in Fig.\,\ref{fig:Mass-rate-S2}.
``S1 SRZ PET'' and ``BS SRZ PET'' curves are always keeping at
the same level during the simulation. However the ``S1 SRZ PP''
curve is always above the ``BS SRZ PP'' curve, which indicates,
that in the S1 case more PP particles are lifted together with PET
particles. In the last 15\,s the mass flow rate of PP particles reaching
the SRZ increases in the simulation without the vertical walls (``S1
SRZ PP'' curve), therefore the mass flow rate of PP particles entering
the FRZ decreases (``S1 FRZ PP'' curve). From the performed simulation
it can be concluded, that the use of the separating vertical walls
helps to reduce the pollution of the discharged PET particles with
PP particles, which makes the use of the separating vertical walls
a preferable design solution.

\subsubsection{Influence of the rotational velocity of the drum}

\begin{figure}
\begin{centering}
\includegraphics[width=7.5cm]{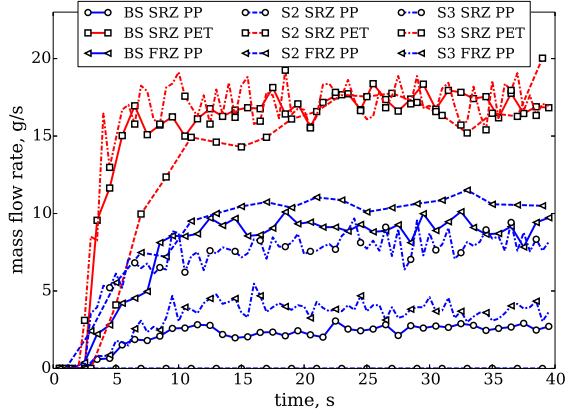}
\par\end{centering}
\caption{Influence of the rotational velocity of the drum: the mass flow rate
of particle removal \label{fig:Mass-rate-SB-S3-S4}}

\end{figure}

A numerical analysis of the influence of the rotational velocity of
the drum is carried out. In addition to the BS simulation, where the
drum rotated at $0.5\pi\,\mathrm{rad/s}$ velocity, two simulations,
where the drum rotates at $0.25\pi\,\mathrm{rad/s}$ (simulation S2
in Table \ref{tab:SensParam}) and at $1.0\pi\,\mathrm{rad/s}$ (S3),
are performed. The mass flow rate of the resulting particle removal
is shown in Fig.\,\ref{fig:Mass-rate-SB-S3-S4}. In all three simulations
the mass flow rate of PET particles removed through the SRZ is at
the same level, however in the S2 case (lowest rotational velocity)
this level is reached a bit later than in the BS and S3 simulations.
The ``BS SRZ PP'', ``S2 SRZ PP'' and ``S3 SRZ PP'' curves show
how many PP particles were lifted and removed together with PET particles
into the SRZ. In the S3 case more PP particles are trapped together
with PET particles (``S3 SRZ PP'' curve) than flow-out through the
opening together with the water (``S3 FRZ PP'' curve). A completely
different situation can be seen in the S2 simulation: just very few
PP particles are lifted into the SRZ and therefore the ``S2 SRZ PP''
curve always remains at zero level. This is observed because, when
the drum rotates more slowly, the PP and PET particles have more time
to separate from each other and PP particles are not caught together
with PET particles. The ``BS SRZ PP'' curve remains at an intermediate
level between corresponding S2 and S3 results. It can be concluded,
that the rotational drum velocity has a big influence on the resulting
purity of the separated particles.

\subsubsection{Influence of the number of lifters}

\begin{figure}
\begin{centering}
\includegraphics[width=7.5cm]{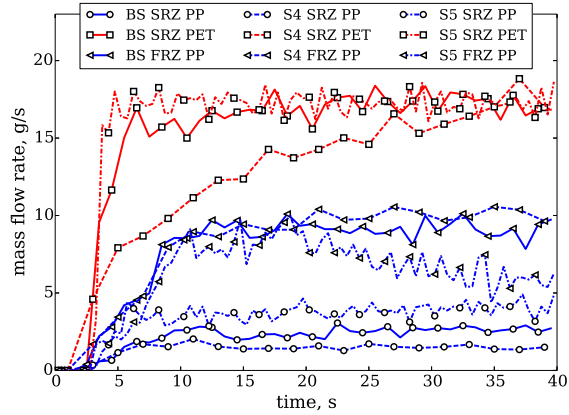}
\par\end{centering}
\caption{Influence of the number of lifters in the drum: the mass rate of particle
removal\label{fig:Mass-rate-SB-S5-S6}}

\end{figure}

In the BS simulation 4 lifters (see Fig.\,\ref{fig:DrumSimple})
on the sides of the drum are used. Simulations with 2 (S4) and 8 (S5)
lifters are performed additionally. The results are presented in Fig.\,\ref{fig:Mass-rate-SB-S5-S6}.
``BS SRZ PET'' and ``S5 SRZ PET'' curves are keeping on the same
level, however the corresponding curve from the S4 simulation, where
only 2 lifters are used, is reaching this level only at about 35\,s.
This indicates, that in the S4 simulation more PET particles are remaining
inside of the drum. Comparing the ``BS SRZ PP'', ``S4 SRZ PP''
and ``S5 SRZ PP'' curves is clear, that the lowest mass rate of
PP particles removed together with PET is in the case when 8 lifters
are used. ``S5 FRZ PP'' has a tendency to decrease during the second
part of the simulation, which indicates, that more PP particles are
remaining in the drum.

\subsubsection{Influence of the feed rate of solid particles}

\begin{figure}
\begin{centering}
\includegraphics[width=7.5cm]{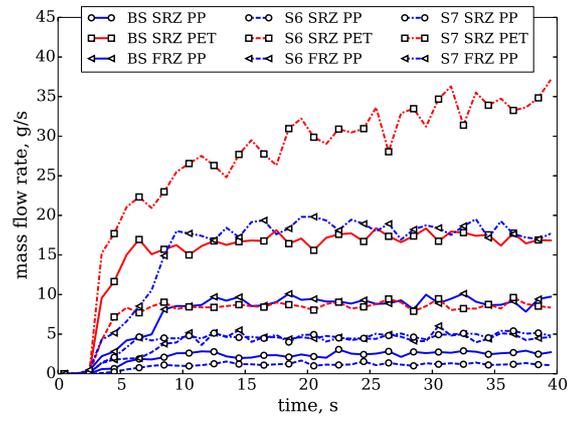}
\par\end{centering}
\caption{The mass flow rate of particle removal: Influence of the feed rate
of solid particles\label{fig:Mass-rate-SB-S7-S8}}

\end{figure}

\begin{figure}
\begin{centering}
\includegraphics[width=7.5cm]{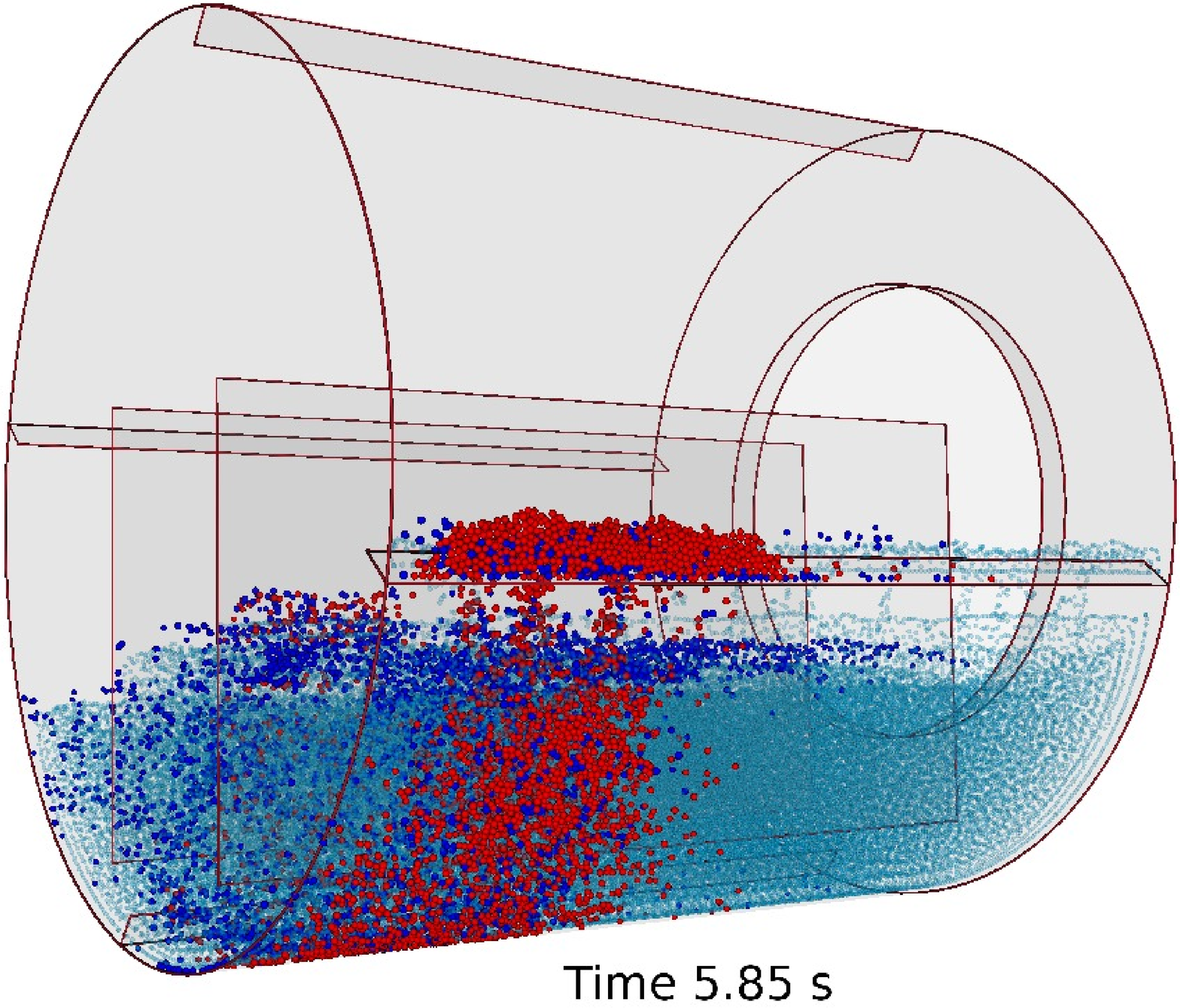}
\includegraphics[width=7.5cm]{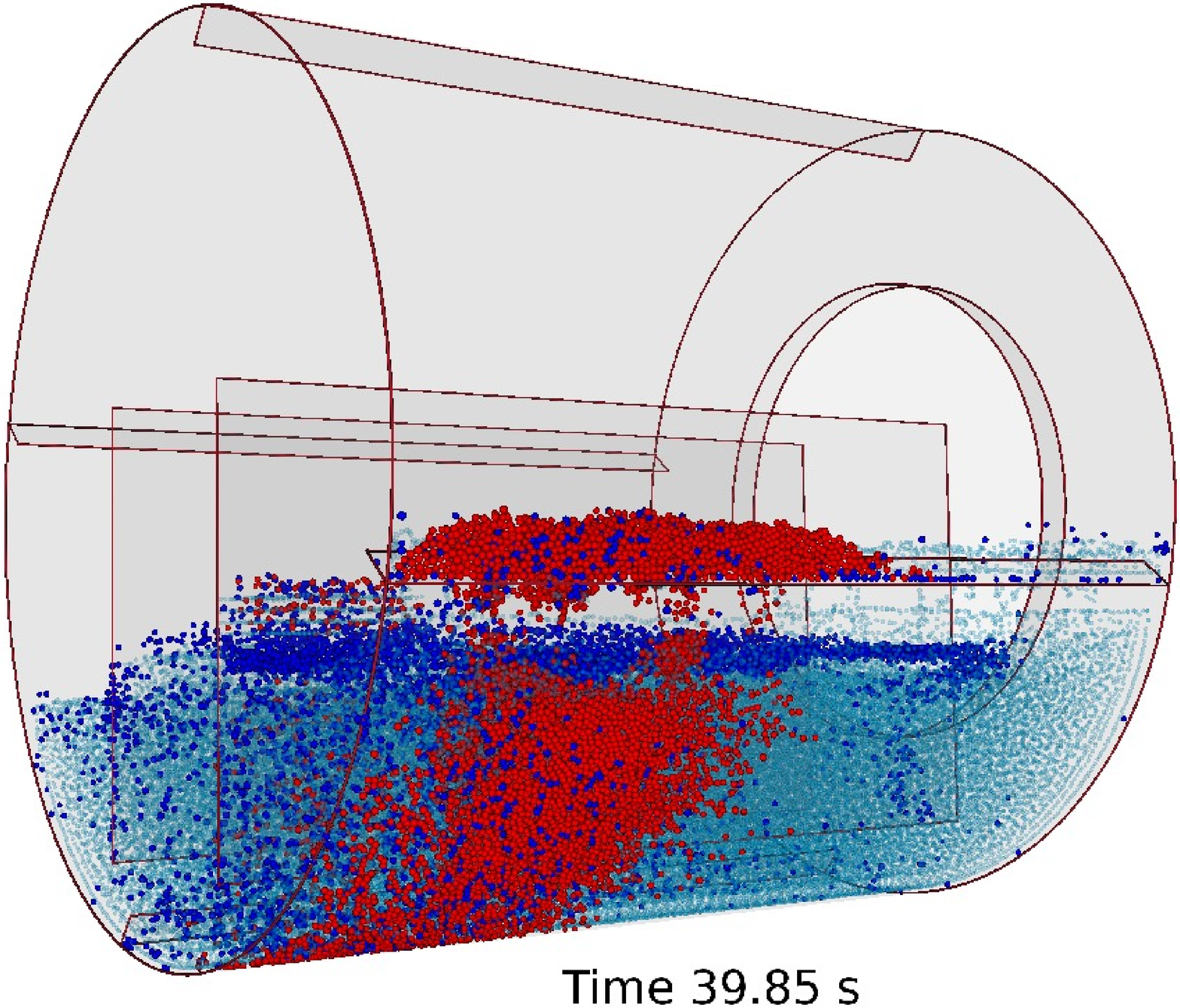}
\par\end{centering}
\caption{Snapshots of the simulation S7\label{fig:Drum_part_S7}}

\end{figure}

The influence of the feed rate of solid particles to the sorting process
is analysed. Simulations with three different feed rates of solid
particles are performed: BS with a feed rate of 29.3\,g/s, S6 with
a two times lower feed rate of 14.65\,g/s and S7 with a two times
higher feed rate of 58.6\,g/s. The results of the simulations are
presented in Fig.\,\ref{fig:Mass-rate-SB-S7-S8}. Comparing the ``BS
FRZ PP'', ``S6 FRZ PP'' and ``S7 FRZ PP'' curves it can be seen,
that the character of these curves is similar, however the level of
the mass flow rate indicates the differences in the feed rate: the
curve ``S6 FRZ PP'' keeps at about 4.5\,g/s, the curve ``BS FRZ
PP'' keeps at about 9.0\,g/s, while the curve ``S7 FRZ PP'' keeps
at about 18.0\,g/s. The influence on the PET particles coming to
the SRZ can be seen from the comparison of the ``BS SRZ PET'', ``S6
SRZ PET'' and ``S7 SRZ PET'' curves. While the ``BS SRZ PET''
and ``S6 SRZ PET'' curves are keeping at a constant level from about
6\,s, the ``S7 SRZ PET'' curve is increasing until the end of the
simulation. This is because the part of the lifter near to the particle
feed zone is fully loaded and therefore cannot lift all sunken PET
particles in simulation S7 as can be seen in Fig.\,\ref{fig:Drum_part_S7}.
With the time the amount of sunken PET particles in the drum is increasing,
and the zone with the sunken PET particles is increasing too. Therefore,
the bigger part of the lifter is used to lift the particles, which
is resulting in the increase of the rate of the removed PET particles
(``S7 SRZ PET'' curve). The differences in the ``BS SRZ PP'',
``S6 SRZ PP'' and ``S7 SRZ PP'' curves indicates, that the amount
of PP particles removed together with the sunken PET particles is
increasing when more particles are fed into the drum. However, the
increase of these removed PP particles corresponds to the increase
of the amount of the removed PET particles.

\subsubsection{Influence of the water feed rate}

\begin{figure}
\begin{centering}
\includegraphics[width=7.5cm]{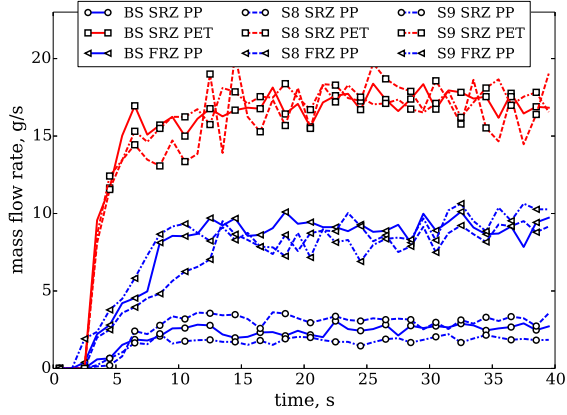}
\par\end{centering}
\caption{The mass flow rate of particle removal: Influence of the feed rate
by solids\label{fig:Mass-rate-SB-S9-S10}}
\end{figure}

The mixture of solid particles can be fed into the drum with different
amounts of water. The influence of the water feed rate is analysed
by performing three simulations with water feed rates 1.035\,l/s,
2.07\,l/s and 4.14\,l/s. The results of these simulations are presented
in Fig.\,\ref{fig:Mass-rate-SB-S9-S10}. The ``BS SRZ PET'', ``S8
SRZ PET'' and ``S9 SRZ PET'' curves are at the same level, which
indicates that the amount of water makes no influence on the settling
and the removal of the PET particles. The small influence of the amount
of the water can be observed by comparing the ``BS SRZ PP'', ``S8
SRZ PP'' and ``S9 SRZ PP'' curves. It can be seen, that the ``S9
SRZ PP'' curve is below the other two curves, which indicates, that
in S9 a bit less PP particles are caught together with the settled
PET particles.

\subsubsection{Comparison of resultant purity of PET}

\begin{figure}
\begin{centering}
\includegraphics[width=7.5cm]{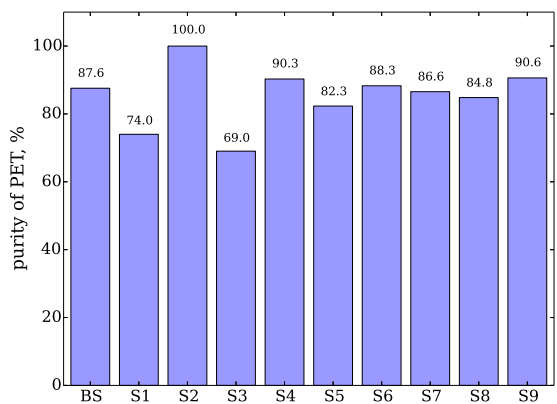}
\par\end{centering}
\caption{Purity of PET \label{fig:Purity-of-PET}}

\end{figure}

The resultant PET purity obtained in the simulations is summarised
in Fig.\,\ref{fig:Purity-of-PET}. Here, the purity is defined as
the mass of PET particles removed in the SRZ divided by the mass of
all particles which were removed in the SRZ:

\[
\mathrm{Purity_{PET}=\frac{\mathrm{mass}_{PET,SRZ}}{\left(\mathrm{mass_{PET,SRZ}+mass_{PP,SRZ}}\right)}}.
\]

As can be seen in Fig.\,\ref{fig:Purity-of-PET}, the resultant purity
for the basic setup (BS) is equal to 87.6\,\%. The removal of the
vertical walls (S1) reduces the purity to 74.0\,\%. With the lowering
of the rotational velocity of the drum the purity of PET increases
up to 100.0\,\%, while the increase of rotational velocity reduces
the purity down to 69.0\,\%. The reduction of the number of used
lifters from 4 to 2 (BS and S4 cases), increases the purity to 90.3\,\%,
while the increase of the number of the lifters (8 lifters in S5 case)
decrease the purity to 82.3\,\%. It was found, that in the tested
range the feed rate of solid particles has just a small influence
on the resultant purity (S6 and S7 cases). The purity of PET increased
from 84.8\,\% to 90.6\,\%, when the feed rate by water was increased
from 1.035\,l/s up to 4.14\,l/s (S8 and S9 cases). It could be concluded,
that from all tested parameters, the rotational velocity has the biggest
influence on the resultant purity of PET.

\section{Conclusions}

In the present study, an application of a coupled DEM-SPH scheme for
the analysis of wet plastic particle separation was presented. The
used DEM-SPH scheme was described and dam break tests were performed.
The results were compared with published results found in literature,
which, together with our earlier study {[}\citealp{Markauskas2017}{]},
let us conclude about the validity of the used technique. The numerical
analysis of polyethylene terephthalate (PET) particle separation from
polypropylene (PP) particles in a rotating drum was performed. The
influence of different operational and design parameters, such as
the rotational velocity, the number of lifters, the feed rate etc.,
was analysed. Numerical results show, that the use of the separating
vertical walls, lower rotational velocity, higher number of lifters
and a higher water feed rate increases the purity of the separated
particles. While the used technique was validated by comparing the
simulation results of dam break problem with published experimental
data, a direct comparison of the numerical simulation of the drum
separation process is a desirable and logical next step in the future. 

\section*{Acknowledgements}

This project has received funding from the European Union\textquoteright s
Horizon 2020 research and innovation programme under the Marie Sklodowska-Curie
grant agreement No. 652862.


\bibliographystyle{elsarticle-num}
\addcontentsline{toc}{section}{\refname}\bibliography{Markauskas_Particle-separation}

\end{document}